# Bandgap renormalization in monolayer MoS$_2$ on CsPbBr$_3$ quantum dot via charge transfer at room temperature

*Subash Adhikari, Ji-Hee Kim\*, Bumsub Song, Manh-Ha Doan, Minh Dao Tran, Leyre Gomez, Hyun Kim, Hamza Zad Gul, Ganesh Ghimire, Seok Joon Yun, Tom Gregorkiewicz and Young Hee Lee\**

Dr. S. Adhikari, Prof. J.-H. Kim, B. Song, Dr. M.-H. Doan, Dr. M. D. Tran, Dr. H. Kim, Dr. H. Z. Gul, G. Ghimire, Dr. S. J. Yun, Prof. Y. H. Lee
IBS Center for Integrated Nanostructure Physics, Institute for Basic Science (IBS), Sungkyunkwan University, Suwon 440-746, Republic of Korea
Department of Energy Science, Department of Physics, Sungkyunkwan University, Suwon 440-746, Republic of Korea
E-mail: (J.-H. Kim) kimj@skku.edu, (Y. H. Lee) leeyoung@skku.edu
Dr. L. Gomez, Prof. T. Gregorkiewicz
van der Waals-Zeeman Institute, University of Amsterdam, Science Park 904, 1098 XH Amsterdam, The Netherlands

**Abstract**

Many-body effect and strong Coulomb interaction in monolayer transition metal dichalcogenides lead to shrink the intrinsic bandgap, originating from the renormalization of electrical/optical bandgap, exciton binding energy, and spin-orbit splitting. This renormalization phenomenon has been commonly observed at low temperature and requires high photon excitation density. Here, we present the augmented bandgap renormalization in monolayer MoS$_2$ anchored on CsPbBr$_3$ perovskite quantum dots at room temperature via charge transfer. The amount of electrons significantly transferred from perovskite gives rise to the large plasma screening in MoS$_2$. The bandgap in heterostructure is red-shifted by 84 meV with minimal pump fluence, the highest bandgap renormalization in monolayer MoS$_2$ at room temperature, which saturates with further increase of pump fluence. We further find that the magnitude of bandgap renormalization inversely relates to Thomas-Fermi screening length. This provides plenty of room to explore the bandgap renormalization within existing vast libraries of large bandgap van der Waals heterostructure towards practical devices such as solar cells, photodetectors and light-emitting-diodes.

**1. Introduction**

Reduction of optical and/or electrical bandgap by quantum confinement Stark effect[1-4] and Coulomb screening[5-10] has been the subject of interest in two-dimensional (2D) van der Waals (vdW) materials. For example, the Stark effect reduces the optical bandgap in $MoS_2$ upon application of uniform electric field,[3,4] while the Coulomb interaction by dielectric/plasma screening modifies the overall band structure of 2D semiconductors.[7,11] This leads to optical and/or electrical bandgap renormalization (BGR), a carrier density-dependent shrinkage of the bandgap.[12-15]

Bandgap modification in 2D semiconductors can also be obtained using mechanical strain,[16-20] which is not highly desirable owing to the strain sensitivity.[21,22] Meanwhile, Coulomb engineering by dielectric/plasma screening using either dielectric materials (2D and bulk) or free carriers from external sources (charge doping, photoexcitation and electric field) offers a non-perturbative and nanoscale control of electrical and optical bandgaps,[9,11,23-28] which is advantageous over the strain. In particular, monolayer $WS_2$ on hexagonal boron nitride (hBN) under large electron doping reveals the spin-orbit splitting, resulting in the BGR of ~325 meV,[26] while the dielectric screening with graphene on monolayer $WS_2$ reduces electronic bandgap by more than 100 meV.[24] However, these phenomena are observed only at low temperature while room temperature phenomena requires high photoexcitation fluence/density.[25] Similarly, the optical BGR in monolayer $MoS_2$ under high photoexcitation at low temperature is shifted by 30-50 meV[23] and the bandgap in $MoS_2$/hBN is reduced from 400 meV at femtosecond time scale to <100 meV in picoseconds time range.[11]

Coulomb engineering with dielectric patterning thus allows for the selective reduction of the bandgap by renormalization phenomenon at nanoscale resolution. More importantly, BGR can offer several possibilities in wide bandgap vdW materials for the development of near-infrared (NIR) optical/opto-electronic devices. However, the electrical/optical BGR shift with dielectric/plasma screening either requires an optical excitation source with high photoexcitation density and/or are observed mostly at low temperature (summarized in SI Table T1). The high photoexcitation also produces structural/phase transition as well as higher order exciton interaction in 2D materials which limits the practical applications of BGR phenomenon for large bandgap reduction in monolayer vdW materials.[25,29-31] The challenge is to achieve tunable electronic structure modifications of large bandgap vdW materials at room temperature and with a low photoexcitation density to realize a robust BGR shift for the development of transition metal dichalcogenides based electronic/optoelectronic devices.

In this study, we demonstrated BGR at room temperature via interfacial electron transfer in the type-II hybrid structure of $MoS_2$ and all-inorganic $CsPbBr_3$ perovskite quantum dots (PQDs). This was achieved from the shift of the optical bandgap as revealed by optical characterization by linear absorption, photoluminescence (PL), and transient absorption (TA). The efficient charge transfer from PQDs to $MoS_2$, further enhanced by the photoexcitation of highly absorbing and luminescent PQDs, allows for the carrier density in $MoS_2$ as high as 7.7 x $10^{12}$ cm$^{-2}$ giving rise to enhanced plasma screening effect in PQD/$MoS_2$ heterostructure. Our heterostructure charge transfer based renormalization phenomenon resulting in a large BGR shift of 84 meV with a low photon excitation density further demonstrates the feasibility of reducing the optical/electrical bandgap in vdW materials by tailoring the carrier concentration.

## 2. Results and Discussion

Various photo-induced processes in semiconductors are schematically illustrated in Figure 1. The excited photon energy creates electron-hole pairs with strong exciton binding energy ($E_{ex}$) typically in monolayer vdW-layered materials, distinct from a quasiparticle bandgap ($E_{gap}$) (Fig. 1a), which is in general indistinguishable from conventional bulk semiconductors at room temperature. Disorder or exciton delocalization can induce binding energy shift, i.e., the Stokes shift ($\Delta_{Stokes}$) (Fig. 1b). Similarly, an external electric field can also induce charge dipoles that modulate the exciton binding energy, known as the quantum-confined Stark effect ($\Delta_{Stark}$). This effect is particularly prominent in quantum-confined structures (Fig. 1c). Further, high carrier injection and/or photoexcitation may produce heavily degenerated situation when the Fermi level moves into the conduction or valence band. In this case, the band filling effect can induce many-body phenomena that renormalize the band structure, resulting in BGR.

All inorganic perovskite crystals with high air stability, large absorption coefficient, and high quantum efficiency can be synthesized with varying chemical compositions and size yielding different bandgaps.[32-37] Among these, pure (20 mg/mL) $CsPbBr_3$ PQDs synthesized by wet-chemistry methods (see Methods for details of synthesis procedure) show a sharp excitonic feature of absorption at 2.426 eV (Fig. 2a) and a narrow PL peak at 2.412 eV, producing a Stokes shift of ~14 meV. This direct bandgap feature in PL with an full width at half maximum (FWHM) of ~80 meV has been observed for PQDs with size distribution of 11±2 nm, as shown in the TEM image together with electron diffraction patterns in the inset.[32,35,36] Figure 2b shows two absorption peaks, A at 1.922 eV and B-excitons at 2.050 eV in a large-area monolayer $MoS_2$ film synthesized by chemical vapor deposition (CVD) method and transferred onto a quartz substrate (see Methods for details of the growth and transfer process). The PL for the same $MoS_2$ film excited at 2.331 eV (532 nm) is dominated by the neutral exciton ($A^0$) peak at 1.929 eV.

The PQD/$MoS_2$ hybrid structure (deposited by spin-coating) (Fig. 2c) shows a dominant PL peak at 1.871 eV, corresponding to a negative trion ($A^-$), with a shoulder emission of $A^0$ at 1.912 eV (see SI Fig. S1 for the details of the PL mapping image and fitting of PL spectra). This redshift of PL peak position together with the shift of spectral weight in the hybrid structure is attributed to the photo-induced electron transfer,[38,39] more importantly from PQDs to MoS$_2$ in our heterostructure.[40] This is further verified by the out-of-plane A$_{1g}$ peak shift (2.47 cm$^{-1}$) in Raman measurement (see SI Fig. S2).[40,41] Moreover, the two exciton absorption peaks (A and B-excitons) of MoS$_2$ at 1.899 and 2.03 eV are redshifted in the hybrid structure by ~23 and ~20 meV, respectively, compared to pristine MoS$_2$ absorption, while the absorption peak of the PQDs in the hybrid structure coincides with that of pristine PQDs (Fig. 2a-c and SI Fig. S3). This absorption shift could originate from either the quantum-confined Stark effect owing to the electric field induced by the interface dipoles in the heterostructure or BGR owing to band filling, as depicted in Fig. 1.

Our hybrid structure of PQD/MoS$_2$, having an electron affinity of 3.8 eV in the PQDs[42,43] and 3.92 eV in MoS$_2$[44-46], constructs the staggered band alignment (type-II) (Fig. 2d). This creates band bending at the PQD/MoS$_2$ interface that allows the transfer of electrons (holes) into MoS$_2$ (PQDs), and is verified by measuring the transfer characteristics of back-gated MoS$_2$ FET (a Hall bar structure shown in the inset) with/without PQDs at $V_{DS}$ = 1 V (Fig. 2e). The pristine MoS$_2$ reveals intrinsic n-type characteristics with a strong gate modulation and a threshold voltage ($V_{th}$) of 4 V from the linear current (see SI Fig. S4a). Meanwhile, the $V_{th}$ is downshifted to -28 V in the PQD/MoS$_2$ device. This large negative $V_{th}$ shift and the increase in $I_{DS}$ by approximately six orders of magnitude implies a more severe n-doping effect in MoS$_2$ owing to the charge transfer at the interface. We estimated the electron concentration ($n$) in PQD/MoS$_2$ using $n = q^{-1}C_g|\Delta V_{th}| = 2.5 \times 10^{12}$ cm$^{-2}$, where $q$ is the electron charge and $C_g$ is the gate oxide capacitance (1.23 × 10$^{-8}$ $F.cm^{-2}$) for 300 nm SiO$_2$, and $\Delta V_{th}$ (-32 V) is the shift in the threshold voltage after electron doping from the PQDs.[47,48] This increase of $n$ in PQD/MoS$_2$ is also measured quantitatively from the Hall bar measurement at a magnetic field of -8 to 8 T (see SI Fig. S4b for details), thereby yielding a similar electron concentration of $\Delta n = 2.4 \times 10^{12}$ cm$^{-2}$. The observed PL shift together with the electrical measurements suggests that in the hybrid structure the electron concentration in MoS$_2$ is increased due to charge transfer induced by band alignment at the interface.

To confirm electron transfer from the PQD to MoS$_2$, we compare the PL intensity mapping (Fig. 3a) of the PQDs on quartz and MoS$_2$ region (dashed blue area) at an excitation energy of 3.062 eV (405 nm) (see SI Fig. S5a,b for PL peak position and FWHM mapping). The PQD PL intensity map clearly demonstrates the PL quenching on MoS$_2$ region. The average PL intensity of PQD on the MoS$_2$ region decreases approximately by 13 times, compared to the PL intensity on the quartz region (Fig. 3b). The quenching of intensity of the PQDs PL in MoS$_2$ region is readily explained by the electron transfer from the PQDs to MoS$_2$.[40] This electron transferred on MoS$_2$ thus produces the redshift as well as intensity quenching in the PL spectra of PQD/MoS$_2$ (Fig. 3b, inset).[38,39] Moreover, no appreciable energy shift nor intensity quenching was observed in a diluted PQD sample (see SI Fig. S6).

Furthermore, the large electron population from the PQDs in the heterostructure by the electron transfer yields A$^-$ with an energy redshift of 50 meV as a consequence of quenched A$^0$ exciton as well as similar B exciton shift of 40 meV (see SI Fig. S7 for details of PL analysis).[39] The time-dependent PL and current-voltage (*I-V*) measurement also rules out the possibility of photo-induced charging effect in our heterostructure, suggesting the presence of both, interfacial charge transfer at dark state and consistent photo-induced charge transfer phenomenon under illumination (see SI Fig. S8). We also verified the photo-induced charge transfer from the PQDs to MoS$_2$ (Fig. 3c) by using time resolved photoluminescence (TRPL) mapping acquired at the PQD PL peak position of 2.412 eV. The fluorescence lifetime of PQDs on MoS$_2$ (dashed blue regions) shows the decrease of lifetime as compared to PQDs on quartz (see SI Fig. S9 for the TRPL spectra and analysis). This is due to the electron transfer from PQDs to MoS$_2$ that decreases the electron density in PQDs thereby reducing the fluorescence lifetime. Hence, both the PL and TRPL measurement confirms the photoinduced charge transfer phenomenon in the heterostructure. However, the significant shift of position and spectral weight in A- as well as B-exciton in steady state absorption (Fig. 2c) and PL measurement (Fig. 2b,c and Fig. 3a,b) in the heterostructure, suggests additional phenomenon apart from interfacial and photoinduced charge transfer from PQD to $MoS_2$.

To elucidate the effective many-body phenomena that result in the large energy shift, we measured spectrally-resolved TA maps for $MoS_2$, PQDs and their heterostructures using a pump energy of 3.543 eV (350 nm) at 22 $\mu J/cm^2$ as a function of probe energy and delay time (Fig. 4a). The TA map of $MoS_2$ (top image) shows photobleaching (PB, negative absorption) of A-exciton at 1.910 eV and B-exciton at 2.058 eV, while that of the PQDs (middle) shows PB at 2.440 eV. The energy of each exciton peak in TA is similar to that in steady-state absorption (Fig. 2b). However, in the PQD/$MoS_2$ heterostructure (bottom image), the A- and B-exciton absorption peaks at an early time delay of 0.3 ps is redshifted ($\Delta E_x$) by 82 meV for A-exciton ($\Delta E_A$, black dashed line) and 78 meV for B-exciton ($\Delta E_B$, gray dashed line) compared to $MoS_2$ throughout the probe delay range. Here, both $MoS_2$ and PQD/$MoS_2$ absorption at a probe delay of 0.2 to 100 ps shows a small blue shift of 10-15 meV due to exciton-phonon scattering[46] (see SI Fig. S10). Similarly, the photo-induced absorption (PIA) peaks in the PQD/$MoS_2$ heterostructure is also red-shifted by 52 meV compared to pristine $MoS_2$ ($\Delta IA$, marked by dashed green lines), while it has a small redshift of 17 meV for PQD on $MoS_2$ (red dashed line). The redshift of absorption processes in both A- and B-exciton and the corresponding shift induced by this in the PIA band suggests the renormalization of the whole band structure of $MoS_2$. This absorption peak shift originates from BGR mediated by the band filling effect in $MoS_2$ through a photo-excited charge transfer from the PQDs as well as pump-induced photoexcitation in $MoS_2$.

Because the band filling also depends on the photoexcited carrier density[23], we further analyze our BGR observation by measuring the fluence-dependent TA spectra for $MoS_2$, PQDs and their heterostructure from 6 to 165 $\mu J/cm^2$ (see SI Fig. S11). The pump fluence was limited under the linear regime of $MoS_2$ bleach intensity to avoid possible many-body interactions originating mainly from large photo-excited carrier density. To compare the exciton energy position of the heterostructure with that of MoS$_2$, the obtained $\Delta E_x$ values are plotted in Fig. 4b. Here the redshift in spectral position of the PB and PIA peaks in the heterostructure were obtained by taking into account the spectral composition and relative change in spectral weight due to interfacial and photoexcited charge transfer (see SI S11). As shown in Fig. 4b, A and B-excitons at the lowest fluence of 6 µJ/cm$^2$ show an energy redshift of 79 meV and 71 meV, respectively. With elevated pump fluence, the $\Delta E_x$ for A-exciton continues to increase up to 84 meV at a maximum fluence of 165 µJ/cm$^2$, and then stabilizes at higher fluence. In contrast, the $\Delta E_x$ for B-exciton increases up to 78 meV at 22 µJ/cm$^2$, and subsequently reduces down to 66 meV at 165 µJ/cm$^2$. Together with the shifts of A and B-exciton, PIA follows the $\Delta E_x$ trend of A-exciton, however with smaller shift (47 meV at the lowest fluence saturating at 63 meV at 80 µJ/cm$^2$). This suggests that high fluence influences both the exciton binding energy and quasiparticle bandgap, as schematically shown in Fig. 4c. Similarly, the carriers accumulated by high fluence further to B-excitons produces plasma screening in A-excitons, consequently modulating the energy redshift or BGR. We note that such a tenable energy shift in the B-excitons at high fluence is not highly appreciable in the A-excitons in our system.

The energy redshift owing to BGR in the PQD/MoS$_2$ heterostructure can also be correlated with the PL. Figure. 5a shows the PL spectra of MoS$_2$ at different pump power using an excitation source of 3.062 eV (405 nm) (see SI Fig. S12 for further details of PL). At the lowest power of 10 µW, the PL spectrum of MoS$_2$ shows that A-exciton is composed of dominant neutral excitons (A$^0_{MoS2}$) at 1.918 eV and a negative trion (A$^-_{MoS2}$) at 1.881 eV owing to intrinsic n-doping in MoS$_2$. This PL spectra at high power (~96 µW) is primarily dominated by negative trions at 1.870 eV, which is attributed to the increased photon carrier density. In contrast, PQD/MoS$_2$ at 10 µW shows the red-shifted PL spectra dominated entirely by the negative trion (A$^-_{PQD/MoS2}$) at 1.852 eV (Fig. 5b), largely by the photoexcited charge transfer from the PQDs as discussed before. At a higher excitation power, $A^-_{PQD/MoS2}$ is slightly redshifted further by maintaining a similar spectral feature at 1.841 eV at ~96 µW. Similarly, the PL intensity of B-exciton becomes prominent gradually with increasing excitation power and the energy redshift.

The energy redshift of A-exciton ($\Delta A = A^0_{MoS2} - A^-_{PQD/MoS2}$) and similarly that of $\Delta B$ exciton from the PL spectra are extracted along with $\Delta E_x$ from TA at the similar excitation energy and power range (Fig. 5c and see SI Fig. S13 and S14 for details of TA). Both $\Delta A$ and $\Delta B$ in the PL and TA demonstrate the increasing $\Delta E_x$ with the power that is nevertheless faintly larger for TA. This is ascribed to the additional photo-induced plasma effect in TA, producing $\Delta A$ of 73 meV for TA as compared to 69 meV in PL at ~80 µW. The exciton energy redshift by photo-induced charge transfer from PQDs was also analyzed using a laser excitation source of 2.331 eV (532 nm), producing A-exciton renormalization in PQD/MoS$_2$ by 61 meV as well as absorption spectrum producing $\Delta A$ of 18 meV (see SI Fig. S15-17 for details of PL, TA, and linear absorption analysis). This confirms the interfacial charge transfer from PQD to MoS$_2$ in the heterostructure, consequently producing the charge-transfer-mediated optical BGR in PQD/MoS$_2$ heterostructure.

The charge-transfer-mediated exciton energy level shift in PQD/MoS$_2$ heterostructure were further studied by scanning tunneling microscopy/spectroscopy (STM/S; see Methods for details regarding the measurements). The Fermi level from STS was shifted by ~36 meV in MoS$_2$ after PQD deposition (see SI Fig. S18 and S19), which confirms the interfacial charge transfer from PQD to MoS$_2$ in the heterostructure. This charge transfer thus lowers the conduction band minimum in PQD/MoS$_2$ by charge screening (or band filling) effect, giving rise to the electronic bandgap renormalization in PQD/MoS$_2$ heterostructure.[9,10,12,27] More importantly, the decrease of conduction band minimum in STS (~36 meV) is consistent with the interfacial charge-transfer-mediated optical BGR of exciton energy level shift ($\Delta A = 18 \pm$ 10 meV) in linear absorption (SI Fig. S17). This further confirms the optical energy level shift and exciton binding energy reduction originating from electrical BGR.

We finally explain the plasma screening effect described with BGR in terms of the Thomas-Fermi (TF) screening length. The transfer characteristics of the MoS$_2$ FET (Fig. 6a) were measured for the photocurrent in MoS$_2$ and PQD/MoS$_2$ heterostructure in the dark and under an excitation energy of 3.062 eV (405 nm) at 134 µW (see Methods and SI Fig. S20 for details of the photocurrent measurement). The MoS$_2$ FET in dark reveals an n-type behavior with a high on/off ratio of $10^6$ and $V_{th}$ of ~11 V from the linear current (see SI Fig. S21). With the addition of PQDs, the drain current ($I_D$) increases and shifts the $V_{th}$ to around -23 V owing to the charge transfer from the PQDs (observed before in Fig. 1e). The MoS$_2$ FET under laser illumination shows a further increase in n-doping effect with the reduced on/off ratio (~$10^5$) by photo-induced electrons and $V_{th}$ at around -31 V. Meanwhile, the on/off ratio of the illuminated PQD/MoS$_2$ heterostructure decreases to $10^1$ with a $V_{th}$ of about -89 V. This high n-doping in the heterostructure originates from the additional charge transfer from the photoexcited electron from the PQDs to MoS$_2$.

From the shift of $V_{th}$ values from the MoS$_2$ position, we also calculated the 2D electron density and hence the TF screening length using $n = q^{-1}C_g|\Delta V_{th}|$ and $L_{TF} = (\varepsilon K_B T/e^2 N)^{1/2}$, where $\varepsilon = \varepsilon_r\varepsilon_0$ is the dielectric constant, $K_B$ is the Boltzmann constant, $T$ is the temperature, $e$ is the electronic charge, and $N$ is $n/t$, with $t$ being the thickness of monolayer MoS$_2$ (Fig. 6b and see SI S22 for details).[47-49] Here, the decrease of TF screening length with electron density is attributed to the plasma screening effect in PQD/MoS$_2$ heterostructure.[9,11,25,28] Similarly Fig. 6c displays the energy redshift of A-exciton (ΔA), plotted against the electron concentration obtained from Fig 6a. The values of ΔA for PQD/MoS$_2$/dark (interfacial charge transfer) was obtained from linear absorption mapping in SI S17g and S17h and for MoS$_2$/light and PQD/MoS$_2$/light (photo-excited charge transfer) from PL shift in SI S12e with 3.062 eV (405 nm) excitation energy at 134 µW. The obtained ΔA is well fitted with carrier density as $\sim n^{4/3}$ from a model.[10] In addition, the excitonic bandgap renormalization of $18 \pm 10$ meV for PQD/MoS$_2$ at dark (Fig. 6c) due to interfacial charge transfer is consistent with the electrical bandgap renormalization of ~36 meV obtained from STM/S analysis (blue star in Fig. 6c and see details in SI Fig. S18d and S19b and S19e).

We correlate the optical bandgap renormalization with TF screening length in Fig. 6d. Our charge-transfer-mediated optical BGR is inversely related to TF screening length ($\Delta A \propto L_{TF}^{-8/3}$) as suggested by $L_{TF} \propto n^{-1/2}$ and $\Delta A \propto n^{4/3}$. The optical BGR induced by interfacial charge transfer in the heterostructure and further enhanced by photoexcited charge transfer in PQD/MoS$_2$ is ascribed to the enhanced plasma screening effect. This diminishes the mean free path of electrons in PQD/MoS$_2$ as well as the coulomb interaction of charges in the heterostructure. This leads to conclude that the magnitude of optical bandgap renormalization can be tailored by carrier concentration in van der Waals heterostructures.

## 3. Conclusions

We have demonstrated the BGR phenomenon in the perovskite/MoS$_2$ heterostructure mediated by a charge transfer. CsPbBr$_3$ perovskite quantum dots can induce electron doping on MoS$_2$ with a type-II band alignment. The photoexcitation of the PQDs and MoS$_2$ together with the electron doping by charge transfer produces the plasma screening effect, consequently giving rise to BGR in MoS$_2$. The increased electron concentration in MoS$_2$ in a heterostructure, with photoexcitation of PQD reveals a large BGR of 84 meV compared to the BGR of 18 meV by a simple charge transfer in the heterostructure. The observed BGR in MoS$_2$ by charge-transfer offers various possibilities for reducing the large bandgap of existing vast libraries in van der Waals materials for optoelectronics devices including transistors, solar cells and photodetectors.

## 4. Experimental Section

**Growth of PQD.** Cesium lead bromide perovskites ($CsPbBr_3$) were synthesized by following the methods introduced by Protesescu et al.[35,50] In this method, 40 mL of ODE and 700 mg of $PbBr_2$ were dried for 1 h at 120 °C under $N_2$ atmosphere. After water removal, 5 mL of dried OLA and 5 mL of dried OA were added to a reaction flask and the temperature was increased to the desired value. After the complete solvation of the $PbBr_2$ salt, 4 mL of a previously synthesized Cs-oleate solution in the ODE was injected. A few seconds later, the nano-crystal solution was quickly cooled down with as ice bath. The product was purified by centrifugation and re-dispersed in hexane. The perovskite synthesis was carried out at 160°C, yielding nano-crystal $CsPbBr_3$ as described in the literature.[35] The concentration of the samples was 20 mg/mL, as determined from the dry weight.

**Sample preparation.** Large area monolayer $MoS_2$ was grown using a promoter-assisted method via chemical vapor deposition.[51] As-grown $MoS_2$ film on silicon wafer was spin-coated with PMMA C4 at 3000 rpm for 60 s. The sample was subsequently dried in an oven (80°C) for 5 minutes and floated on a deionized water on a Petri dish. The PMMA/$MoS_2$ film floating on water was rinsed thrice and transferred onto quartz and $SiO_2$/Si substrates. The PMMA film was subsequently washed with acetone, isopropanol alcohol, and ethanol solution followed by annealing the sample at 350°C under Ar/$H_2$ environment to completely remove the organic contaminations on the $MoS_2$ film. For the electrical and hall measurement, the samples transferred onto $SiO_2$/Si were patterned by photolithography process for etching and electrode deposition followed by Cr/Au (10/50 nm) deposition using electron-beam evaporator at high vacuum. Similarly for the STM/S measurements, $MoS_2$ films were transferred onto p-doped Si after removing the $SiO_2$ layer by hydro fluoric acid (HF) treatment. As prepared PQD on hexane was spin coated at 1000 rpm for one minute in all $MoS_2$ films under different substrates. The sample was than dried in vacuum desiccator for ~24 hours.

**Characterization.** The linear absorption of the pristine and heterostructure samples were analyzed using the Jasco V-670 spectrophotometer. Confocal absorption spectra imaging using a lab-made laser confocal microscopy system with a tungsten-halogen lamp was used to obtain absorption mapping and local spectra at a resolution of 1 μm diameter, detected by spectrometer. Raman, PL, and TRPL spectra were obtained under ambient conditions using a commercial multifunctional optical microscopy system (NT MDT, NTEGRA Spectra PNL) using two different excitation source of 2.331 eV (532 nm) and 3.062 eV (405 nm). The spatial resolution of the system was indicated to be ~380 nm with an objective lens having numerical aperture of 0.7. TRPL measurement was set up with a time-correlated single photon counting system with a 3.062 eV (405 nm) pulsed laser with a repetition rate of 50 MHz and an pulse width down to 60 ps. Hall bar measurements were performed in a physical property measurement system (Quantum Design Inc.) coupled with the Agilent B1500A semiconductor device analyzer. TA measurements were performed using a 1-kHz Ti: sapphire regenerative amplifier (Libra, Coherent) operating at 790 nm which was divided into two beams. One operates an optical parametric amplifiers (TOPAS prime, Coherent) generating ultraviolet to mid-infrared range of laser light, which was used as a tunable pump pulse from visible to near-infrared (1.24-3.1 eV) with a pulse duration of 200 fs. The other was focused onto a nonlinear crystal to generate a white light continuum and used as a probe pulse that was detected by using a 256-pixel InGaAs sensor (HELIOS, Ultrafast systems). Photocurrent measurements were conducted at high vacuum ($10^{-6}$ Torr) with a Keithley (4200-SCS) parameter analyzer in the dark and laser illumination condition. STM/S measurements were conducted for pristine and PQD/MoS$_2$ heterostructure at room temperature using Omnicron VT-STM. Prior to STM measurements, both MoS$_2$/Si and PQD/MoS$_2$/Si were cleaned in a UHV chamber with a base pressure < 1.0 x $10^{-10}$ T by heating them at < 100 °C for 2 hrs. Electrochemically etched W tips were used after electron bombardment to remove the surface oxide and contaminants. The d*I*/d*V* spectra were recorded by using a lock-in amplifier with a modulation voltage of 50 mV at 919 Hz. All the optical and electrical measurements were carried out by calibrating the instruments with standard references. For the instrument calibration and fitting of absorption and photoluminescence spectra, Gauss and Lorentz function were used within an accuracy of 15% error. The error bars in the optical and electrical measurements represent the errors estimated from the fitting or from data averaging process.

**Supporting Information**

Supplementary information detailing Table T1 and Figures S1-S22 containing summary of bandgap renormalization in 2D materials, PL, TRPL, Raman, linear absorption, *I-V* characteristics, Hall bar analysis, TA analysis, photo-current measurement, STM/S analysis of monolayer $MoS_2$ and $PQD/MoS_2$ heterostructure. This material is available from the Wiley Online Library or from the author.

**Acknowledgements**

This work was supported by the Institute for Basic Science (IBS, R011-D1).

Received: ((will be filled in by the editorial staff))
Revised: ((will be filled in by the editorial staff))
Published online: ((will be filled in by the editorial staff))

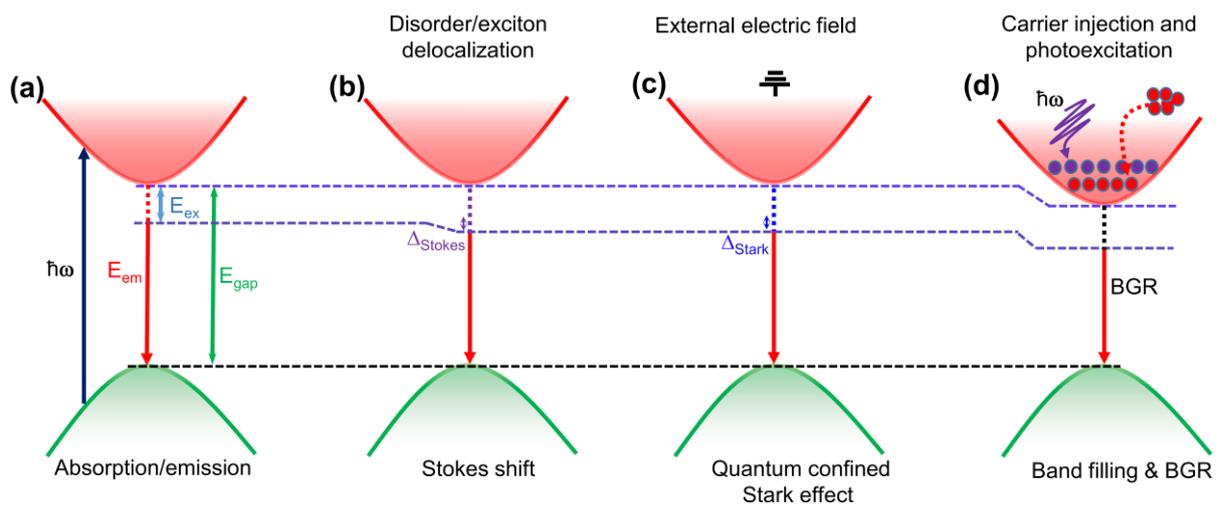

**Figure 1.** Schematic illustration of emission shift in semiconductor: a) Optical excitation followed by emission via the exciton energy level in semiconductor. b) Stokes shift invoked from disorder or exciton delocalization. c) External electric field (or source-drain voltage)

yields quantum confined Stark effect. d) High carrier injection and/or optical excitation generate band filling and BGR.

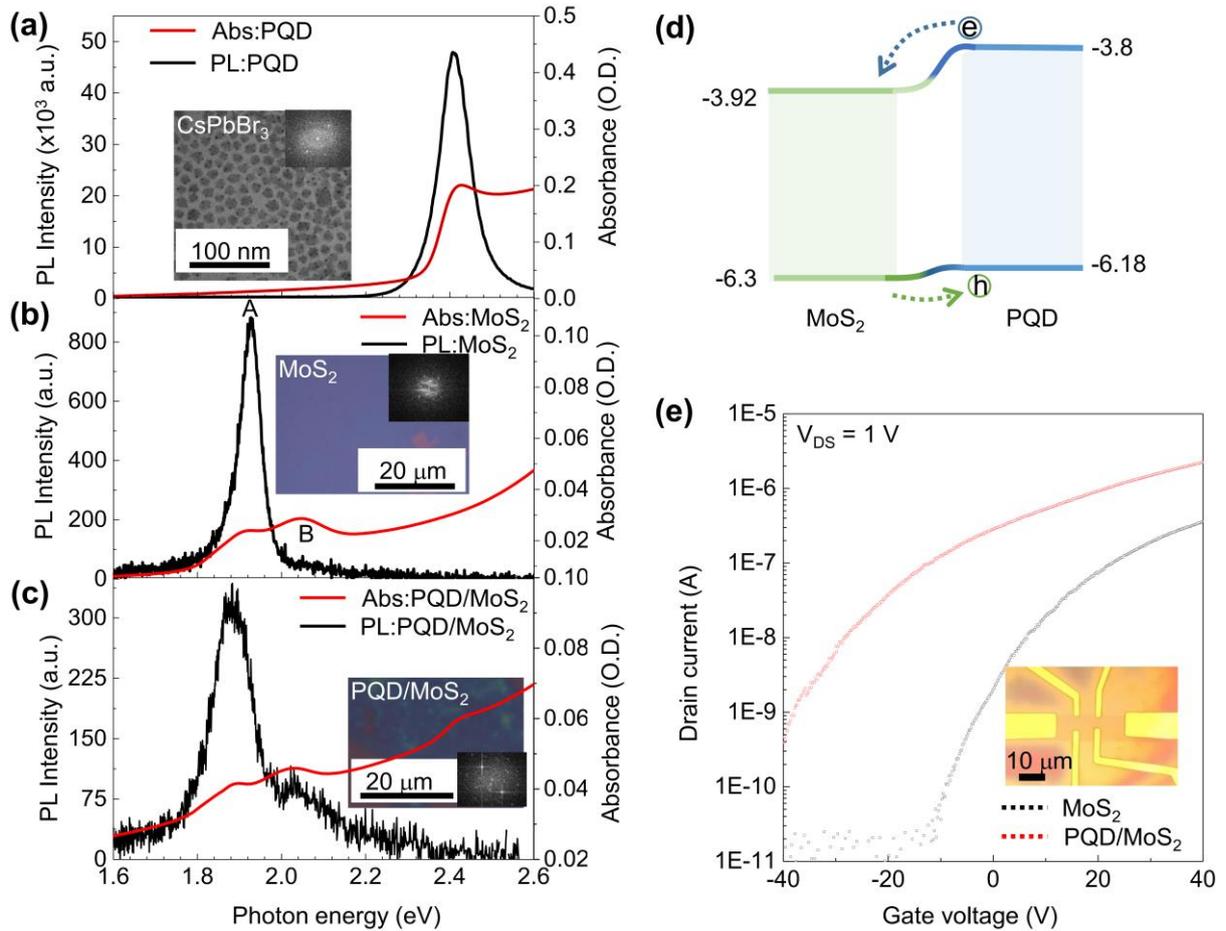

**Figure 2.** Optical and electrical characteristics of MoS$_2$, PQD and their heterostructure: a) Absorbance and PL of PQDs. Excitation energy for PL was 3.062 eV (405 nm). Inset shows TEM image of PQDs with electron diffraction pattern. b), c) Similar spectra as (a) for monolayer MoS$_2$ and PQD/MoS$_2$ heterostructure with an excitation energy of 2.331 eV (532 nm). Inset shows the optical image and the electron diffraction pattern of PQD, MoS$_2$ and PQD/MoS$_2$ heterostructure. The electron diffraction pattern shows the individual PQD and MoS$_2$ crystal together with the co-existing PQD and MoS$_2$ in PQD/MoS$_2$ heterostructure. d)

Schematic of band alignment and charge transfer in PQD/MoS$_2$ heterostructure. e) Transfer characteristics of MoS$_2$ and PQD/MoS$_2$ at $V_{DS}$ = 1 V. Inset shows the optical image of the fabricated PQD/MoS$_2$ Hall bar structure device.

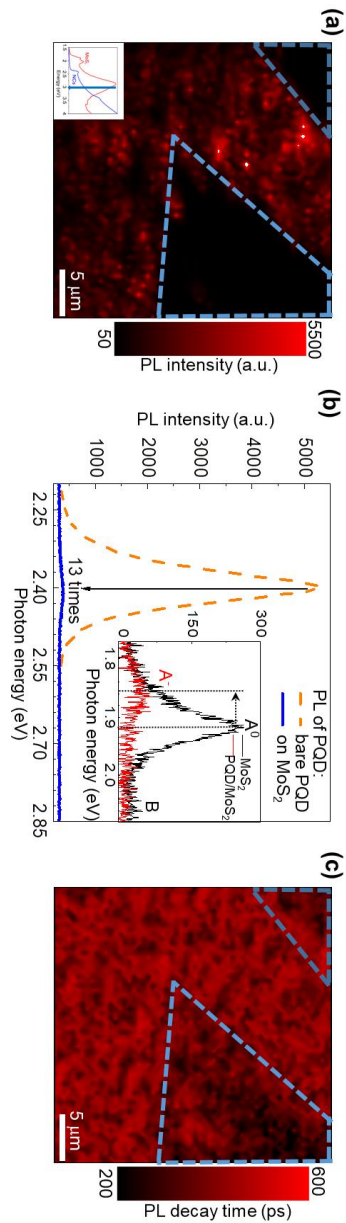

**Figure 3.** PL and charge transfer characteristics of PQD: a) PL intensity mapping of PQDs on quartz and MoS$_2$ (dashed blue line) region. Excitation energy was 3.062 eV (405 nm) in the inset with the light blue arrow in the absorption schematic. b) The PL spectra obtained from the MoS$_2$ and quartz region in (a). Inset shows the PL spectra of MoS$_2$ and PQD/MoS$_2$ under the same excitation source with arrow indicating the amount of A-exciton redshift. c) Fluorescence lifetime image of PQDs on quartz and MoS$_2$ (blue dashed line) region. The color scale bar represents the fluorescence lifetime of PQDs.

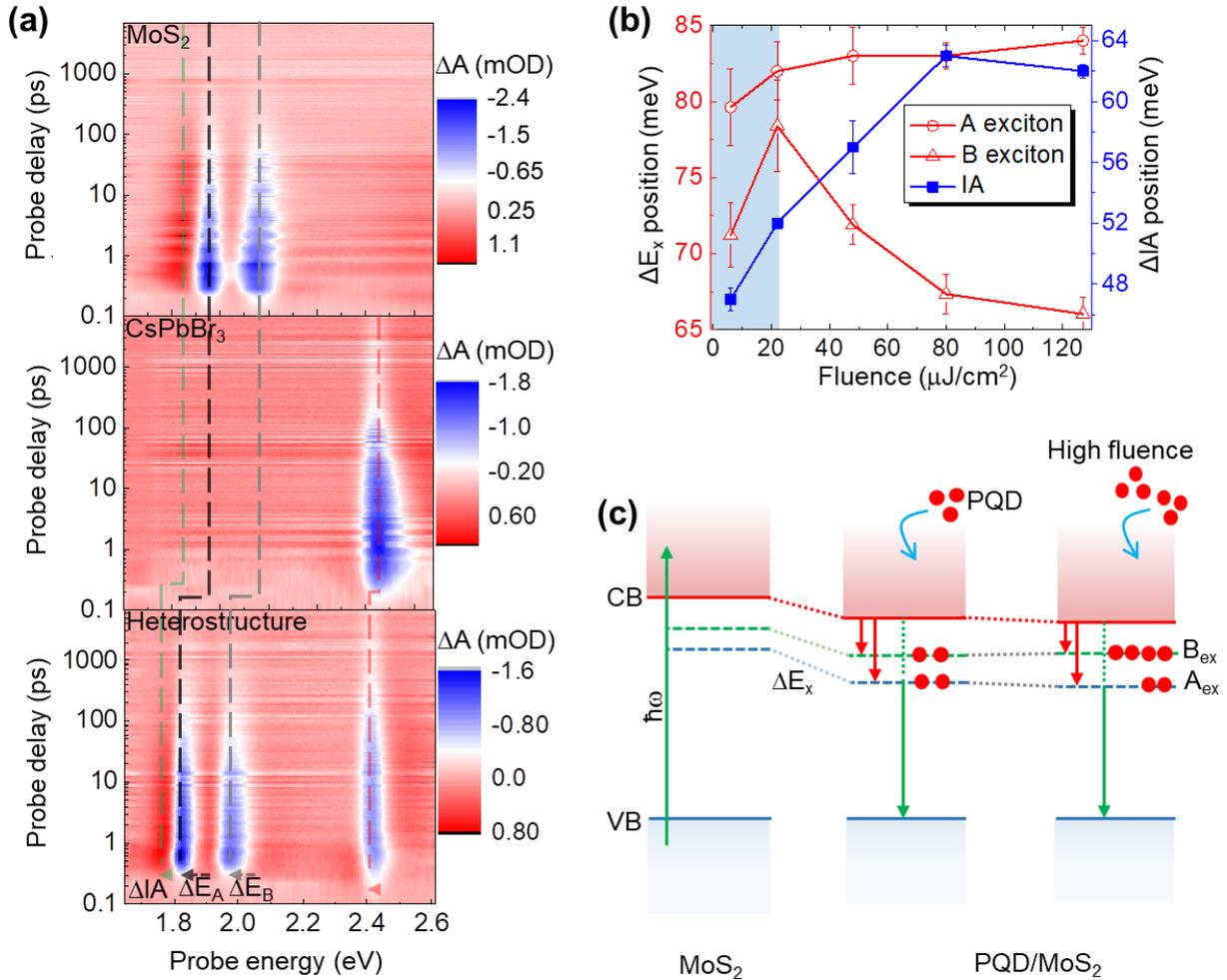

**Figure 4.** TA spectroscopy for BGR analysis in PQD/MoS$_2$ heterostructure: a) Spectrally- and temporally-resolved TA maps for MoS$_2$ (top), PQD (middle) and PQD/MoS$_2$ (bottom). The color scale bar represents the intensity of the A and B-exciton bleach in optical density (mOD). The pump energy was 3.543 eV (350 nm), indicated by the arrow in the inset, with a fluence of 22 µJ/cm$^2$. The dashed lines represent the redshift of induced absorption and exciton energy levels in the heterostructure from their initial positions in MoS$_2$ and PQD that are plotted with varying pump fluences in (b). The light blue region represents the maximum

A- and B-exciton redshifts and the error bars in the shift are obtained from fitting of the TA spectra. c) Schematic illustration of energy shift through various energy states at low and high pump fluences.

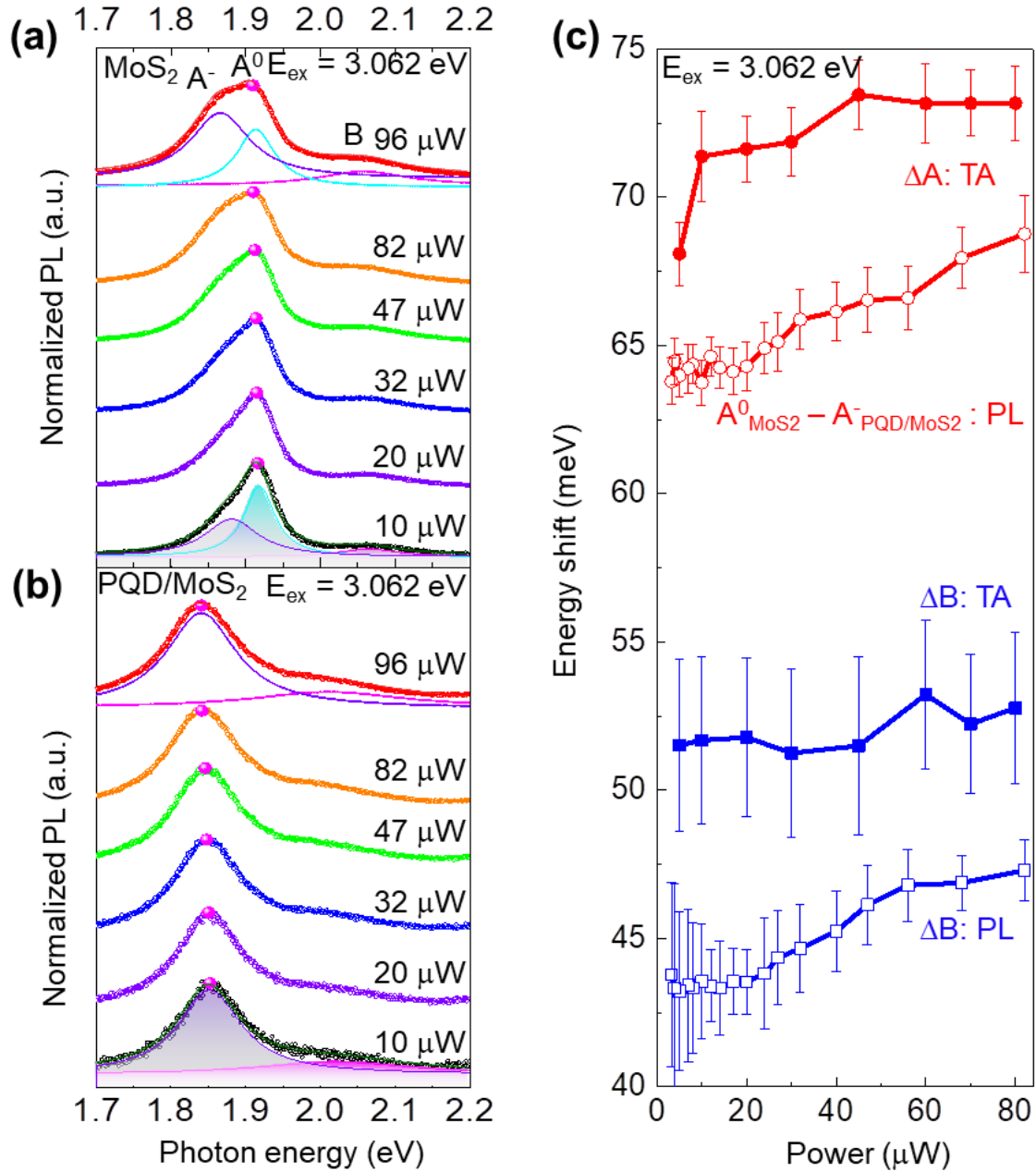

**Figure 5.** Correlation of PL and exciton energy redshift: a), b) Normalized PL of $MoS_2$ (a) and $PQD/MoS_2$ (b) at various laser powers. The solid dots indicate the spectral shift and peak intensity of neutral exciton and negative trions in $MoS_2$ and $PQD/MoS_2$. c) The energy

redshift of A- and B-excitons obtained from PL and TA with 3.062 eV (405 nm) excitation source at various powers. The error bars in the shift are obtained from fitting of the TA spectra.

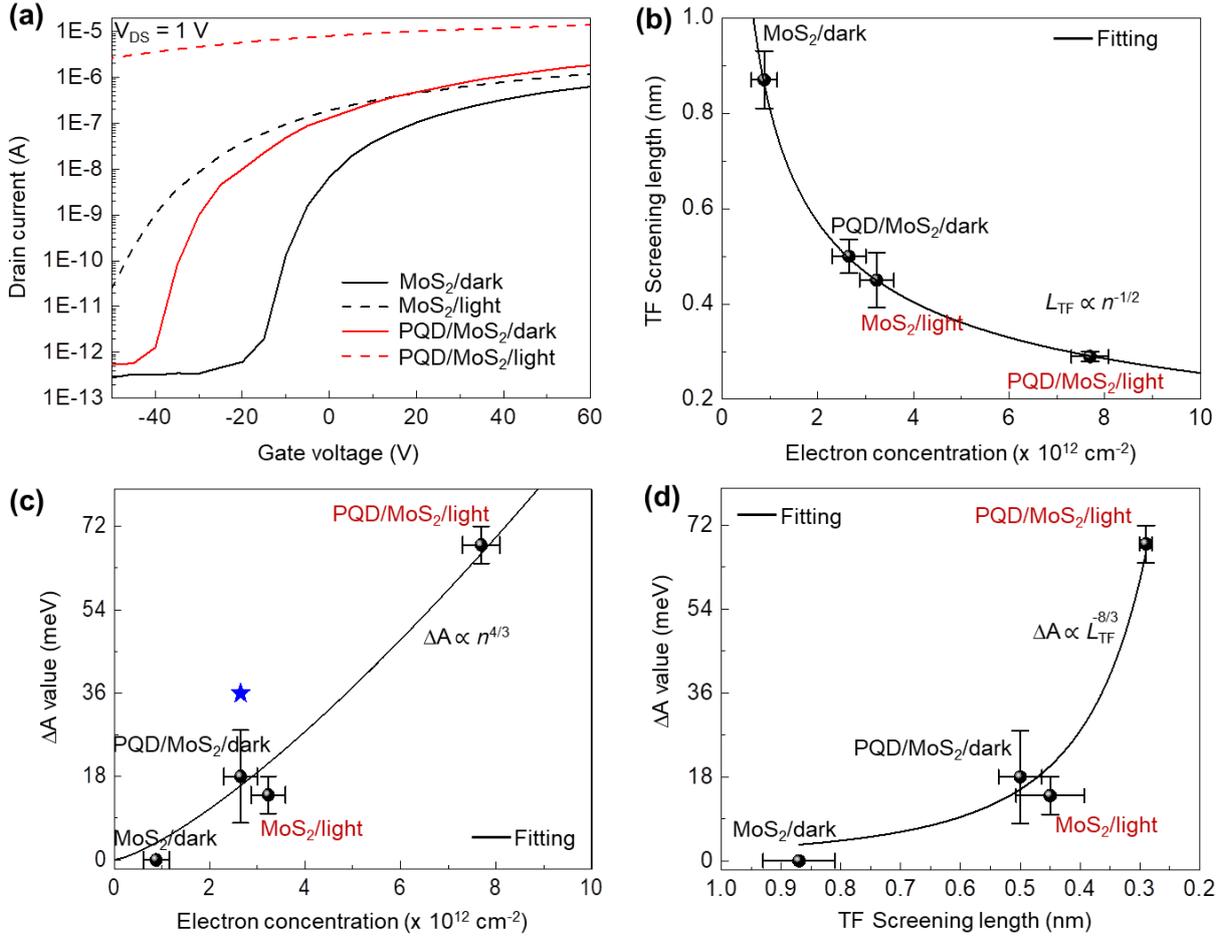

**Figure 6.** Carrier concentration and TF screening length in PQD/MoS$_2$: a) Transfer characteristics of MoS$_2$ with and without PQD at dark and under photoexcitation (E$_{ex}$ = 3.062 eV at 134 µW). PQD/MoS$_2$ under photoexcitation shows shift of threshold voltage ($V_{th}$) as well as increase of current level owing to charge transfer doping. The $V_{th}$ values were used for calculating the electron concentration and thus the Thomas-Fermi (TF) screening length in MoS$_2$ and PQD/MoS$_2$ heterostructure at dark and under photoexcitation in (b). The TF

screening length is inversely proportional to the square root of electron concentration. c) The BGR of A-exciton obtained from the absorption shift (dark) with the electron concentration in MoS$_2$ and PQD/MoS$_2$. The electrical BGR measured from STM/S is marked by green star. d) Correlation between optical BGR and TF screening length.

**The Table of Contents Entry:**

Charge tranfer mediated bandgap renormalization in MoS$_2$ has been demonstrated using highly luminescent CsPbBr$_3$ perovskite quantum dots. The interfacial as well as photoexcited electron doping from quantum dots produces a plasma screening in MoS$_2$ that reduces the optical bandgap in MoS$_2$ by ~84 meV. More importantly, this bandgap renormalization was observed at room temperature and at a low photo-excitation density, further demonstrating the possibilities of inducing large bandgap reduction in van der Waals materials by tailoring the carrier concentration.

**Keywords**
MoS$_2$ monolayer, CsPbBr$_3$ quantum dot, charge transfer, bandgap renormalization, room temperature

*By Subash Adhikari, Ji-Hee Kim\*, Bumsub Song, Manh-Ha Doan, Minh Dao Tran, Leyre Gomez, Hyun Kim, Hamza Zad Gul, Ganesh Ghimire, Seok Joon Yun, Tom Gregorkiewicz and Young Hee Lee\**

**Bandgap renormalization in monolayer MoS$_2$ on CsPbBr$_3$ quantum dot via charge transfer at room temperature**

**ToC Figure**

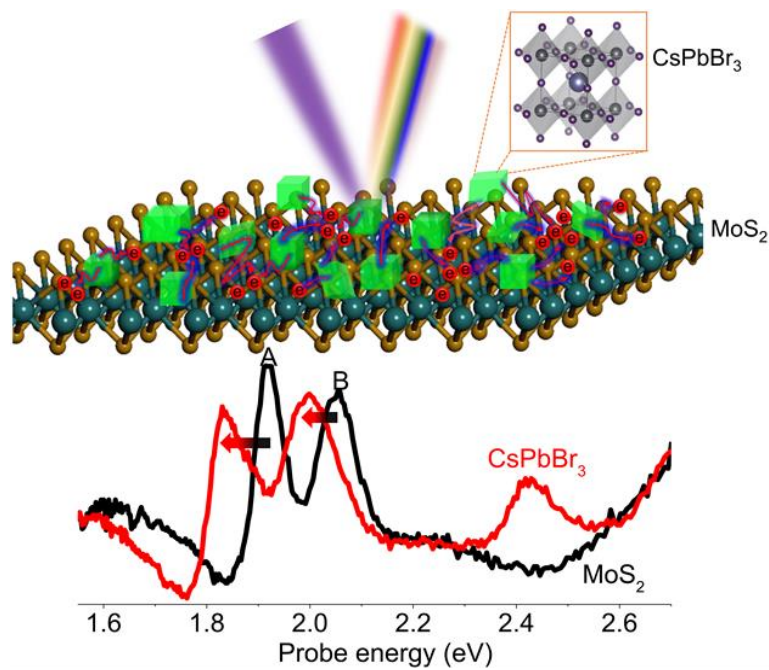

# Supporting Information

**Bandgap renormalization in monolayer MoS₂ on CsPbBr₃ quantum dot via charge transfer at room temperature**

*Subash Adhikari, Ji-Hee Kim\*, Bumsub Song, Manh-Ha Doan, Minh Dao Tran, Leyre Gomez, Hyun Kim, Hamza Zad Gul, Ganesh Ghimire, Seok Joon Yun, Tom Gregorkiewicz and Young Hee Lee\**

**T1. Summary of various optical/electrical bandgap renormalization processes in monolayer 2D transition metal dichalcogenides**

| Sample | BGR shift (meV) | Temperature T (K) | Method | Optical fluence N (cm$^{-2}$) | Drawback |
|---|---|---|---|---|---|
| MoS$_2$ [23] | 30-50 | 10 | TA | 8.2 x 10$^{12}$ | high N, low T |
| MoS$_2$/Graphene [11] | 400-90 | 50 | t-ARPES | 6 x 10$^{12}$ | fs-ps time range, high N, low T |
| MoS$_2$/Pentacene [46] | 6-17 | 300 | TA | low fluence | Unstable pentacene |
| WS$_2$ [25] | 73 | 300 | TA | 3 x 10$^{12}$ | high N |
| K-doped WS$_2$/hBN [26] | 325 | 85 | ARPES | 1.7-2.9 x 10$^{13}$ * | high N, low T |
| Graphene/WSe$_2$ & WS$_2$ [24] | 100-300 | 4-70 | R/white light | | low T |
| MoSe$_2$/hBN/Ru [S1] | 250 | 77 | STM/STS | | low T |

*K-doped density on WS$_2$/hBN

**Table T1.** Electrical and optical bandgap renormalization (BGR) summarized for monolayer 2D transition metal dichalcogenides under various measurement methods including transient absorption (TA), time- and angle-resolved photoemission (tr-ARPES), reflection (R) and scanning tunneling microscopy and spectroscopy (STM and STS) at various measurement conditions. The drawback of these observed BGR phenomenon is also highlighted.

S1.     **PL analysis of MoS$_2$ and PQD/MoS$_2$**

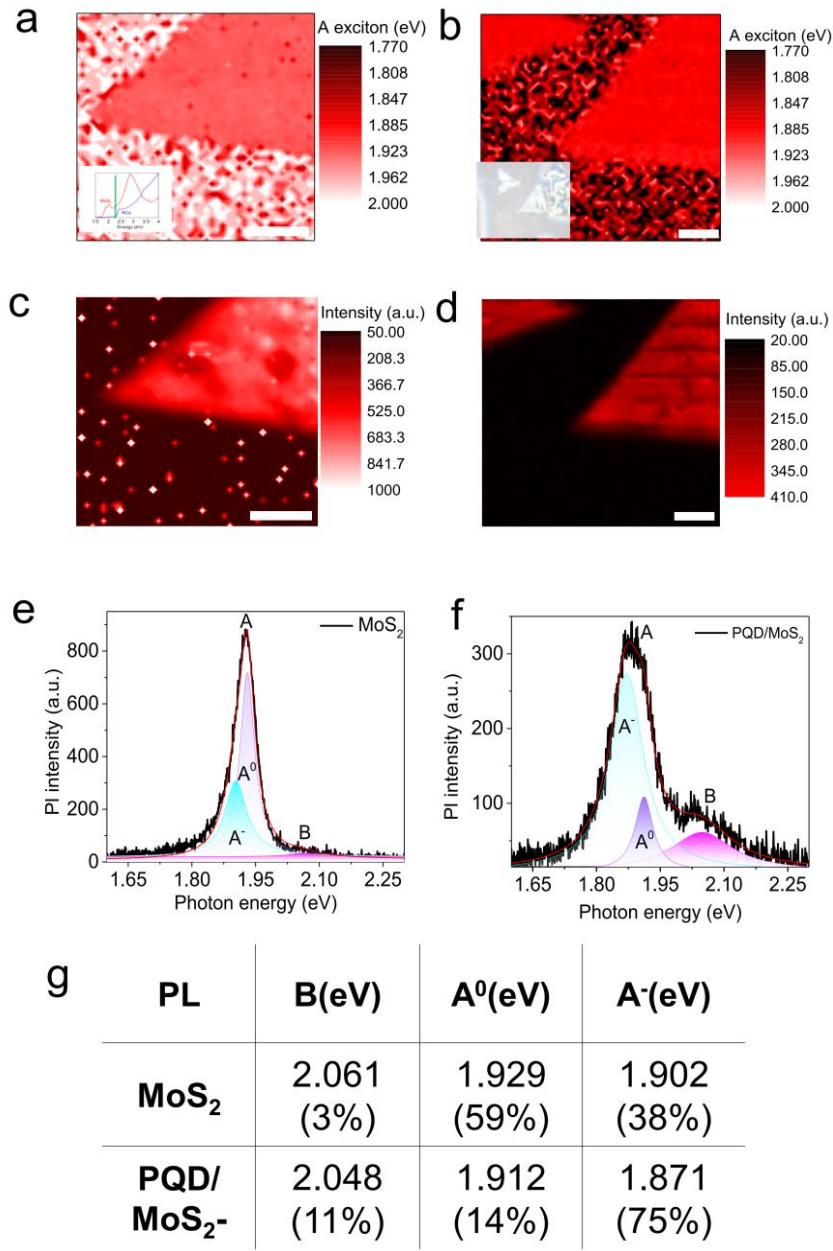

**Figure S1.** PL of $MoS_2$ and $PQD/MoS_2$ exciting only $MoS_2$ excitons. (a, b) Peak position and (c, d) peak intensity map of A exciton of $MoS_2$ and $PQD/MoS_2$. The scale bar indicates 5 μm. (e) PL spectra of $MoS_2$ and (f) $PQD/MoS_2$ fitted with three Lorentzian peaks of neutral exciton ($A^0$), trion ($A^-$) and B exciton. (g) Table showing the position of $A^0$, $A^-$ and B exciton energies of $MoS_2$ and $PQD/MoS_2$ with their respective spectral weight in parentheses. The excitation energy for the mapping was 2.331 eV (532 nm) as shown in inset of (a). The optical image of $MoS_2$ after spin coating PQD is shown in inset of (b).

S2. **Raman analysis of $MoS_2$ and $PQD/MoS_2$**

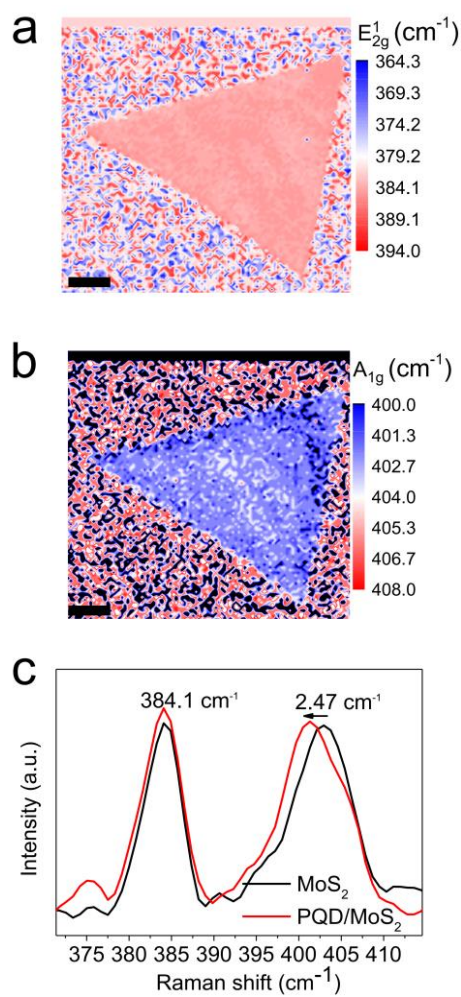

**Figure S2.** Raman spectroscopy of PQD/MoS$_2$. Peak position map of (a) E$^1_{2g}$ and (b) A$_{1g}$. The scale bar indicates 5 µm. (c) Average Raman spectra of pristine MoS$_2$ and PQD/MoS$_2$ showing the shift of A$_{1g}$ peak. The excitation energy for the mapping was 2.331 eV (532 nm). Diluted (30%) PQD are spin coated at 2500 rpm for random dispersion of PQD on MoS$_2$. This was done to avoid the saturation of the PQD PL which otherwise will affect the Raman mode observation.

**S3.** Linear absorption spectra MoS$_2$, PQD and PQD/MoS$_2$

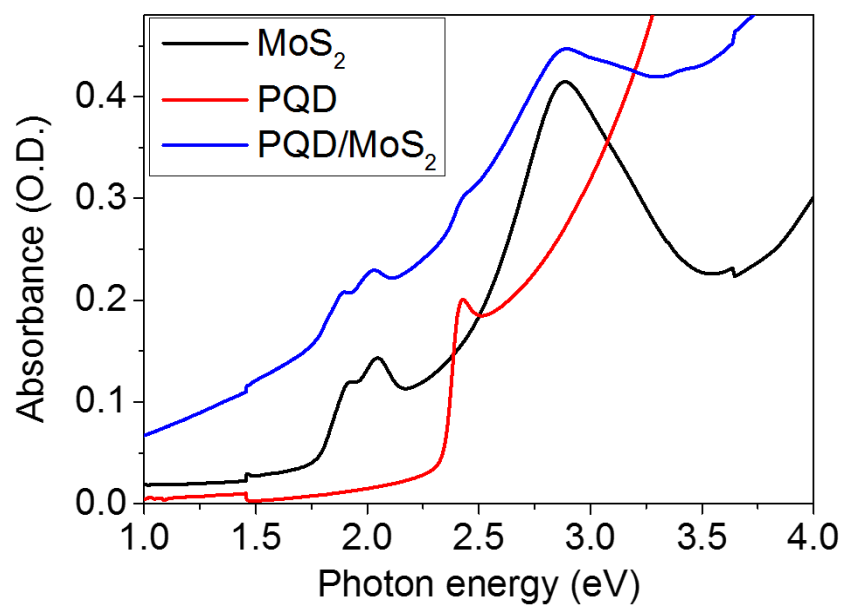

**Figure S3.** Absorption spectra of PQD, MoS$_2$ and the heterostructure of PQD/MoS$_2$. The linear absorption spectra in Figure 2 are plotted together at a larger wavelength window to show the various exciton features in PQD and MoS$_2$.

**S4.** Carrier concentration analysis of MoS$_2$ and PQD/MoS$_2$

Carrier concentrations were obtained via threshold voltage in the transfer characteristics of MoS$_2$ and PQD/MoS$_2$. The linear regions of the current-voltage (*I-V*) curve were extracted to obtain the threshold value ($V_{th}$). The difference of $V_{th}$ in MoS$_2$ and PQD/MoS$_2$ using $n = q^{-1}C_g|\Delta V_{th}|$ gives the excess carrier concentration in PQD/MoS$_2$. This $V_{th}$ was also determined from the difference of the voltage in the $\ln I - V$ curve. The carrier concentration extracted from the gate dependent current-voltage analysis was further confirmed from B-field dependent measurement of MoS$_2$ and PQD/MoS$_2$ (10 x 50 µm$^2$ channel length) in hall bar geometry. The slope of the Hall resistance, $R_{xy} = (\frac{1}{ne})B$, where n is electron concentration, e is charge of electron and B is magnetic field, provides value of the carrier concentration in MoS$_2$ and PQD/MoS$_2$.

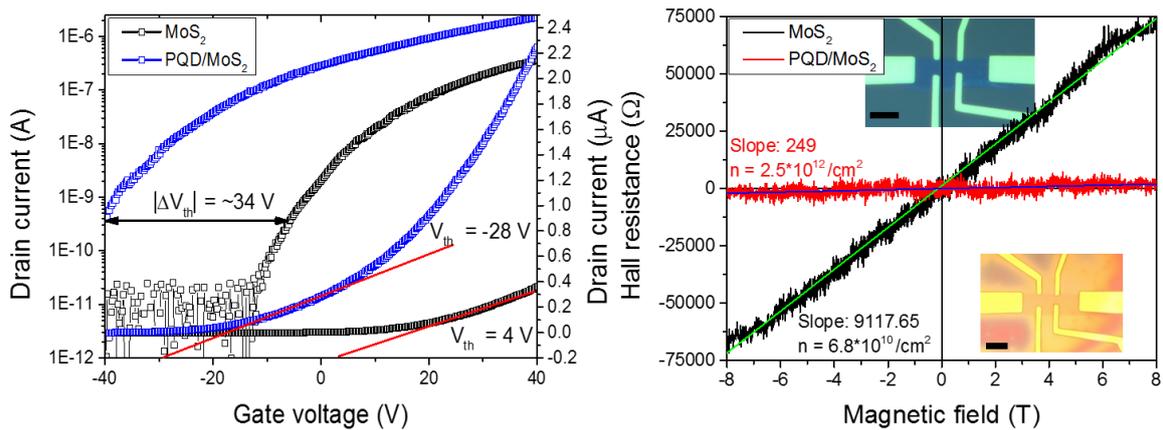

**Figure S4.** Gate dependent current-voltage and Hall bar analysis. (a) Transfer characteristics of MoS$_2$ and PQD/MoS$_2$ at $V_{DS}$ = 1 V plotted in log (left axis) and normal (right axis) scale for measuring the threshold voltage shift ($V_{th}$). (b) Magnetic field dependent Hall resistance measurement of MoS$_2$ and PQD/MoS$_2$ with their respective slope and carrier concentration values. Inset shows the pristine MoS$_2$ (top) and PQD/MoS$_2$ (bottom) hall bar device. The scale bar indicates 10 µm.

**S5.    PL analysis of PQD on MoS$_2$ and only on Quartz substrate**

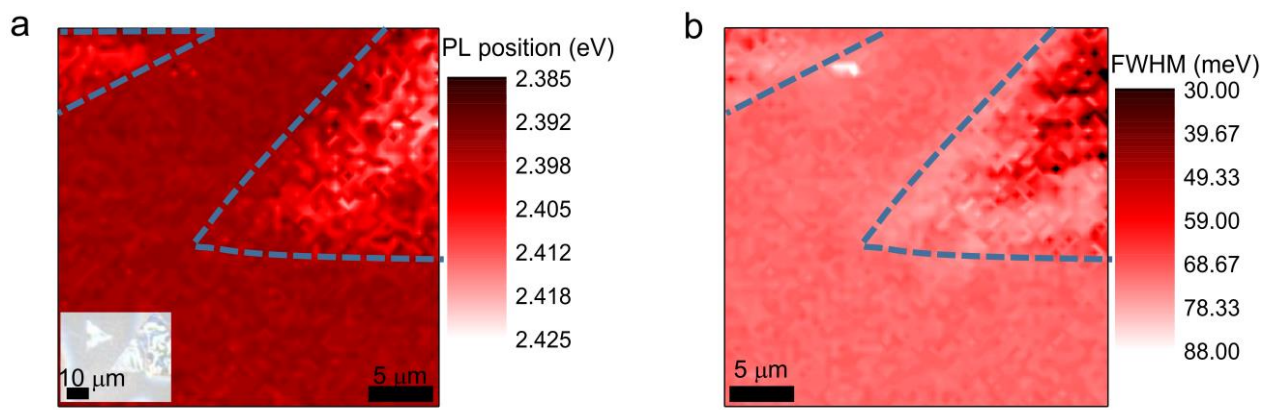

**Figure S5.** PL mapping of PQD peak. (a) PL peak position map of PQD dispersed on MoS$_2$ and on quartz substrate (outside MoS$_2$ region). Inset shows the optical image of the sample used for mapping. The PL peak position of PQD on quartz region shows an average value at 2.399 ± 0.012 eV. (b) FWHM mapping of the PL peak in (a). The blue dashed lines shows the position of PQD dispersed on MoS$_2$ region. The excitation energy for the mapping was 3.062 eV (405 nm). The PL peak position of the PQDs on the MoS$_2$ region blue-shifts by ~20 meV, attributed to the size distribution of PQDs (Ref. 49) while the narrowing of FWHM is due to the charge transfer from PQDs on MoS$_2$ regions.

**S6.    Concentration dependent PQD on MoS$_2$ by PL analysis**

In order to study charge transfer characteristics depending on the PQD concentration the photoluminescence of PQD/MoS$_2$ heterostructure with as-prepared and 50% diluted PQDs were measured. The photoluminescence of MoS$_2$ in the heterostructure with as-prepared PQDs showed a large redshift and intensity quenching compared to pristine MoS$_2$. This is due to the increased charge screening via electron transfer from the ensemble of PQDs (Fig. 1a, inset) to MoS$_2$. However with 50% diluted PQDs, the charge screening in MoS$_2$ is reduced by the isolated nanoparticles, resulting in blueshift of MoS$_2$ photoluminescence compared to MoS$_2$ with as-prepared PQDs. The isolated nanoparticle also reduces the photoluminescence intensity of PQDs on MoS$_2$ due to insufficient PQD contents, resulting into negligible electron doping in MoS$_2$ (similar to pristine MoS$_2$ photoluminescence).

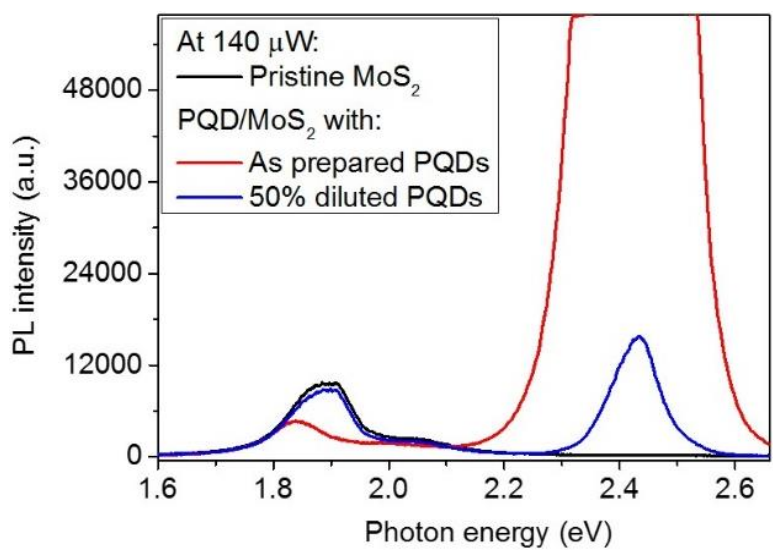

**Figure S6.** PL spectra of pristine MoS$_2$ and PQD/MoS$_2$ heterostructure under as-prepared and 50%-diluted PQD. The redshift and intensity quenching of MoS$_2$ in the heterostructure sample are observed in as-prepared PQDs. However 50%-diluted PQD/MoS$_2$ shows no significant shift in the PL spectra in MoS$_2$, signifying the absence of charge transfer. This is due to the reduced PL intensity in 50%-diluted PQDs that eventually reduces the amount of electron transfer from PQDs to MoS$_2$. The excitation energy for the PL measurement is 405 nm (3.062 eV).

**S7.    PL analysis of MoS$_2$ and PQD/MoS$_2$ with photo-excitation of PQD**

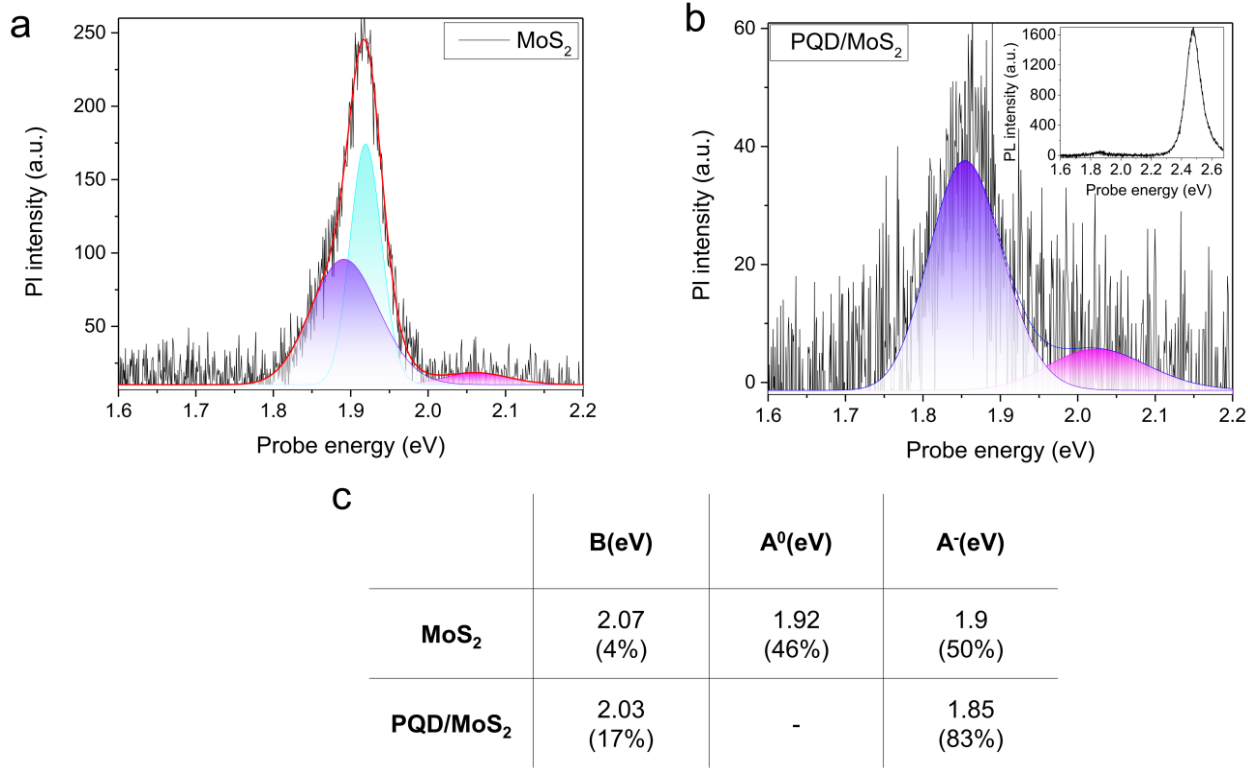

**Figure S7.** PL of MoS$_2$ and PQD/MoS$_2$ exciting both MoS$_2$ and PQD. (a) PL spectra of MoS$_2$ and (b) PQD/MoS$_2$ fitted with Lorentzian peaks. Three Lorentzian peaks of neutral exciton (A$^0$), trion (A$^-$) and B exciton were used for MoS$_2$ fitting while PQD/MoS$_2$ were fitted by trion (A$^-$) and B exciton. Inset in Fig. (b) shows the same spectra in wider energy range with PL peak of MoS$_2$ and PQD. (c) Table showing the position of A$^0$, A$^-$ and B exciton position of MoS$_2$ and PQD/MoS2 with their respective spectral weight in parentheses. The excitation energy used in PL measurement was 3.062 eV (405 nm).

**S8.     Photo-charging phenomenon in PQD/MoS$_2$ heterostructure.**

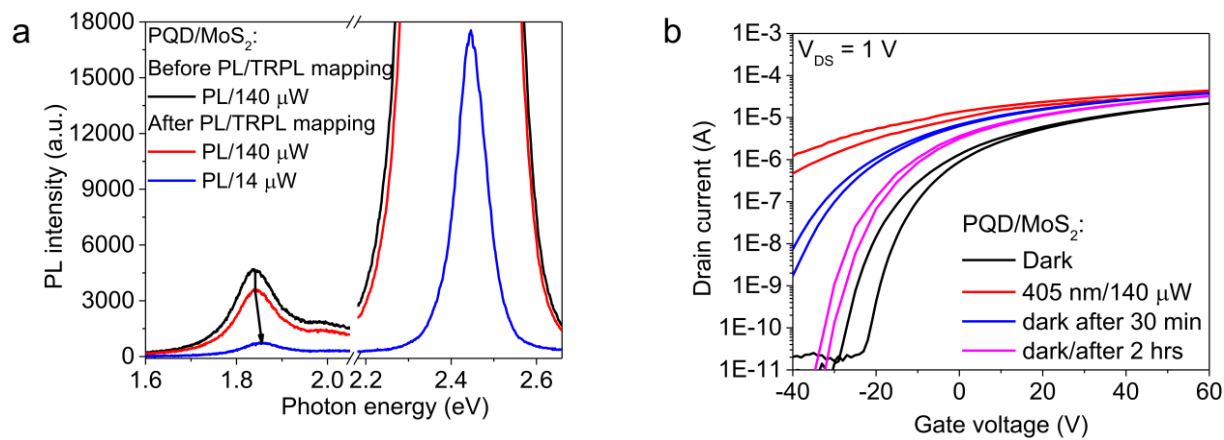

**Figure S8.** Time dependent photo-charging effect. (a) PL spectra of PQD/MoS$_2$ heterostructure before and after PL and consecutive TRPL mapping (Figure 3a,c in main manuscript) at various excitation power. (b) Transfer characteristic of PQD/MoS$_2$ FET at dark and illumination state measured at $V_{DS}$ = 1 V. Both this measurements reveal that there is no significant photo-induced charging effect in the PQD/MoS$_2$ heteorstructure.

**S9. Time-resolved photoluminescence spectra**

The time-resolved photoluminescence (TRPL) spectra obtained from the TRPL mapping data in Fig. 3c also allows us to estimate the fluorescence lifetime of PQD on quartz and MoS$_2$ region as shown in Fig S7a. The TRPL spectra fitted using bi-exponential function, $y = A_1 e^{-x/\tau_1} + A_2 e^{-x/\tau_2}$, yields faster component ($\tau_1$: trap state or Auger recombination) and a slower component ($\tau_2$: radiative recombination, see Fig. S7b).[2,3]

The values of $\tau_1$ and $\tau_2$ listed in the inset table (right) in Fig. S7a shows the decreased lifetime for the PQDs on MoS$_2$ compared to that for PQDs on quartz. This is because of the larger electron density for PQDs on Quartz which decays by radiative process while for PQDs on MoS$_2$ region the electron transfer from PQD to MoS$_2$ region decreases the electron density in PQDs yielding a faster recombination.

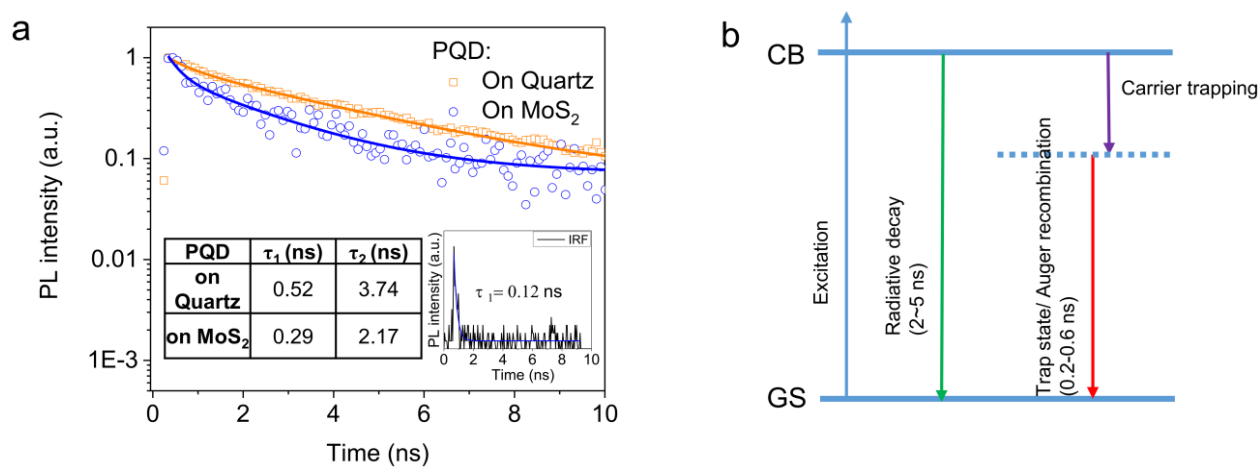

**Figure S9.** TRPL analysis. (a) TRPL spectra of PQD dispersed on Quartz and MoS$_2$ which were acquired at the PL peak position of PQD at 2.412 eV. The inset table shows the decay time of PQD PL while the inset figure shows the instrument response function (IRF). The excitation energy used in TRPL measurement was 3.062 eV (405 nm). (b) Schematic summary of the exciton dynamics of PQD based on the lifetime components extracted from the TRPL spectra fitting.

**S10.  TA spectral feature at pump-probe time delay**

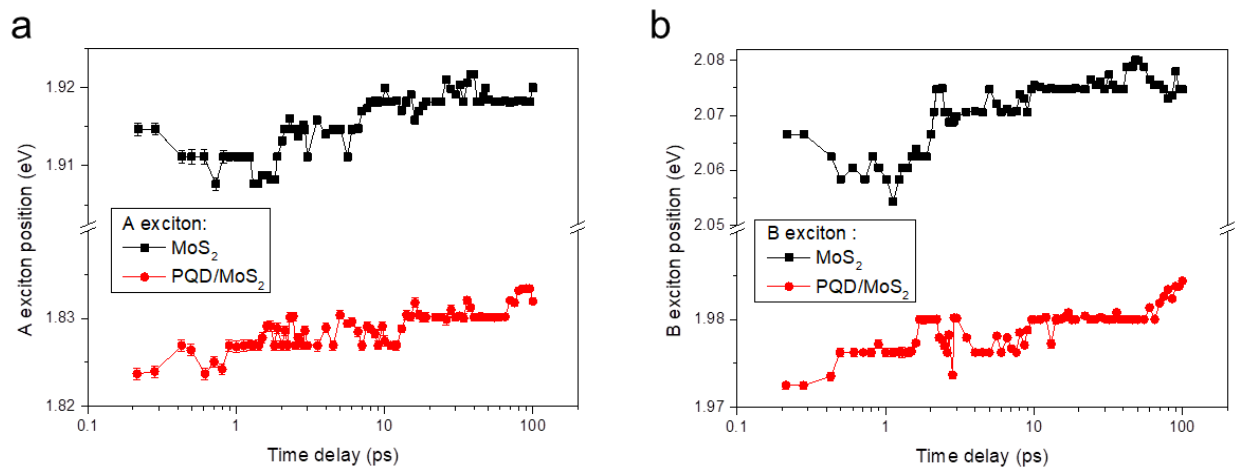

**Figure S10.** Exciton position at various time delay. (a) A exciton position of $MoS_2$ in the pristine and heterostructure sample obtained from the fitting of TA spectra in Figure 4a until 100 ps time delay. (b) Similar spectra in (a) obtained for B exciton. Both A and B exciton in $MoS_2$ and $PQD/MoS_2$ show a blueshift at later time delay (<0.4 ps) compared to its initial position at 0.2 ps. This blueshift arises from the photoexcited carrier decay by inelastic process such as exciton-phonon scattering.[45] The error bars in the exciton shift was obtained from the standard error in the fitting of absorption peaks.

**S11. Transient absorption spectra of $MoS_2$, PQD and $PQD/MoS_2$ with photoexcitation of 3.543 eV (350 nm)**

Based on the previous study by Sie E. J. et. al. (Ref 23 in the main manuscript), the increase in electron density in $MoS_2$ significantly redshifts and broadens the exciton energy levels. This redshift and the broadening of the exciton levels which are manifested as the negative absorption peaks in the TA spectra affect the photo-induced absorption (PIA) bands (positive absorption signal) mainly at the lower energy side of the TA spectra. Hence the free carrier induced shift and the broadening affects the overall TA spectra in the lower energy region while the higher energy PIA bands are not affected significantly. This phenomenon has thus been regarded as bandgap renormalization in $MoS_2$ and many other TMDCs. Similar phenomenon of the spectral redshift of negative absorption and the PIA bands were observed in our pristine and $PQD/MoS_2$ heterostructure (see Fig. S11a). More importantly, the $PQD/MoS_2$ heterostructure in our measurement under various fluence have a larger broadening as compared to pristine $MoS_2$ because of the additional electron transferred due to the interfacial as well as photoexcited electrons in PQD. Hence within the linear range of the TA spectra, we could observe a significant TA shift due to bandgap renormalization in $PQD/MoS_2$ heterostructure. Here it is important to mention that the TA measurements were performed in a large area uniform $MoS_2$ monolayer and the same region under the PQD was used for $PQD/MoS_2$ heterostructure TA measurements.

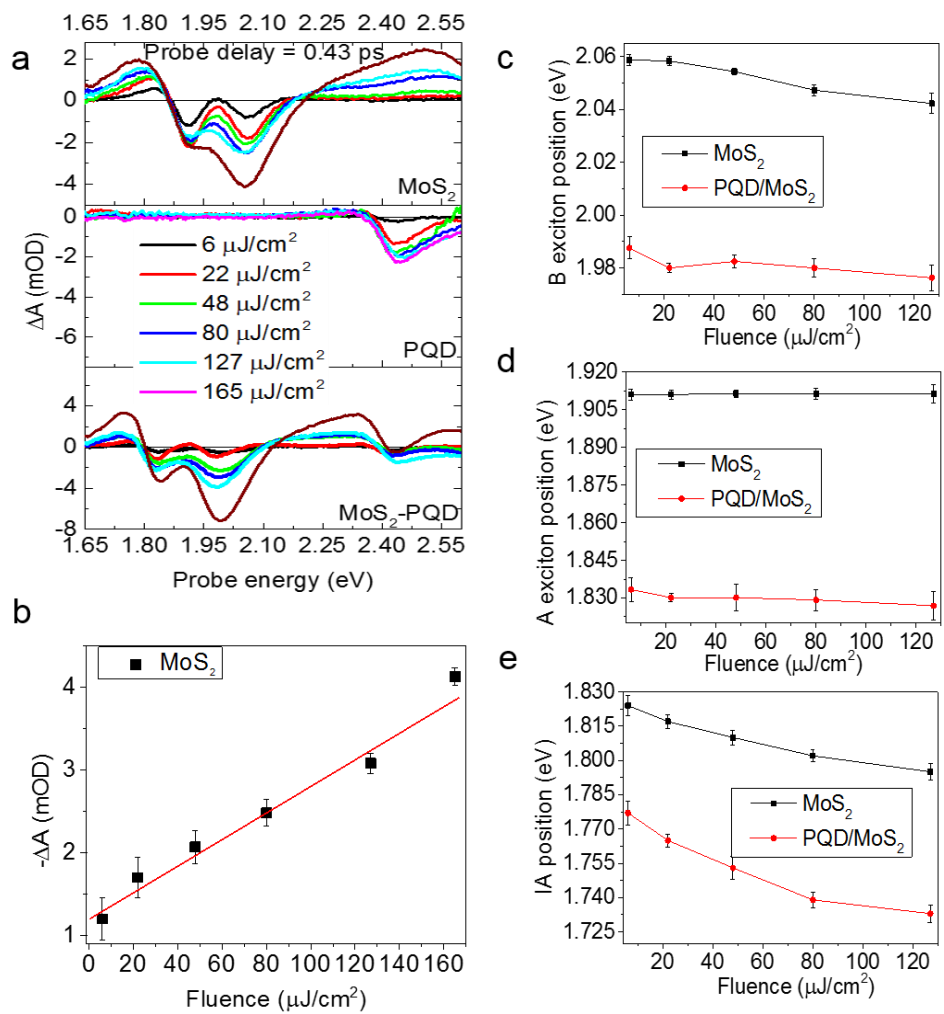

**Figure S11.** TA spectra of MoS$_2$, PQD and heterostructure. (a) Fluence-dependent TA spectra of MoS$_2$, PQD and PQD/MoS$_2$ heterostructure showing the exciton bleach and induced absorption. (b) Plot of bleach intensity at various pump fluence for MoS$_2$. Solid line is a linear fit and the error bars are obtained from multiple spectra scanning. Energy level position of (c) B exciton, (d) A exciton and (e) induced absorption (IA) of MoS$_2$ and PQD/MoS$_2$ obtained from (a). The pump energy used in TA measurement was 3.543 eV (350 nm).

## S12. Power-dependent PL analysis of MoS$_2$ and PQD/MoS$_2$ with photo-excitation of PQD

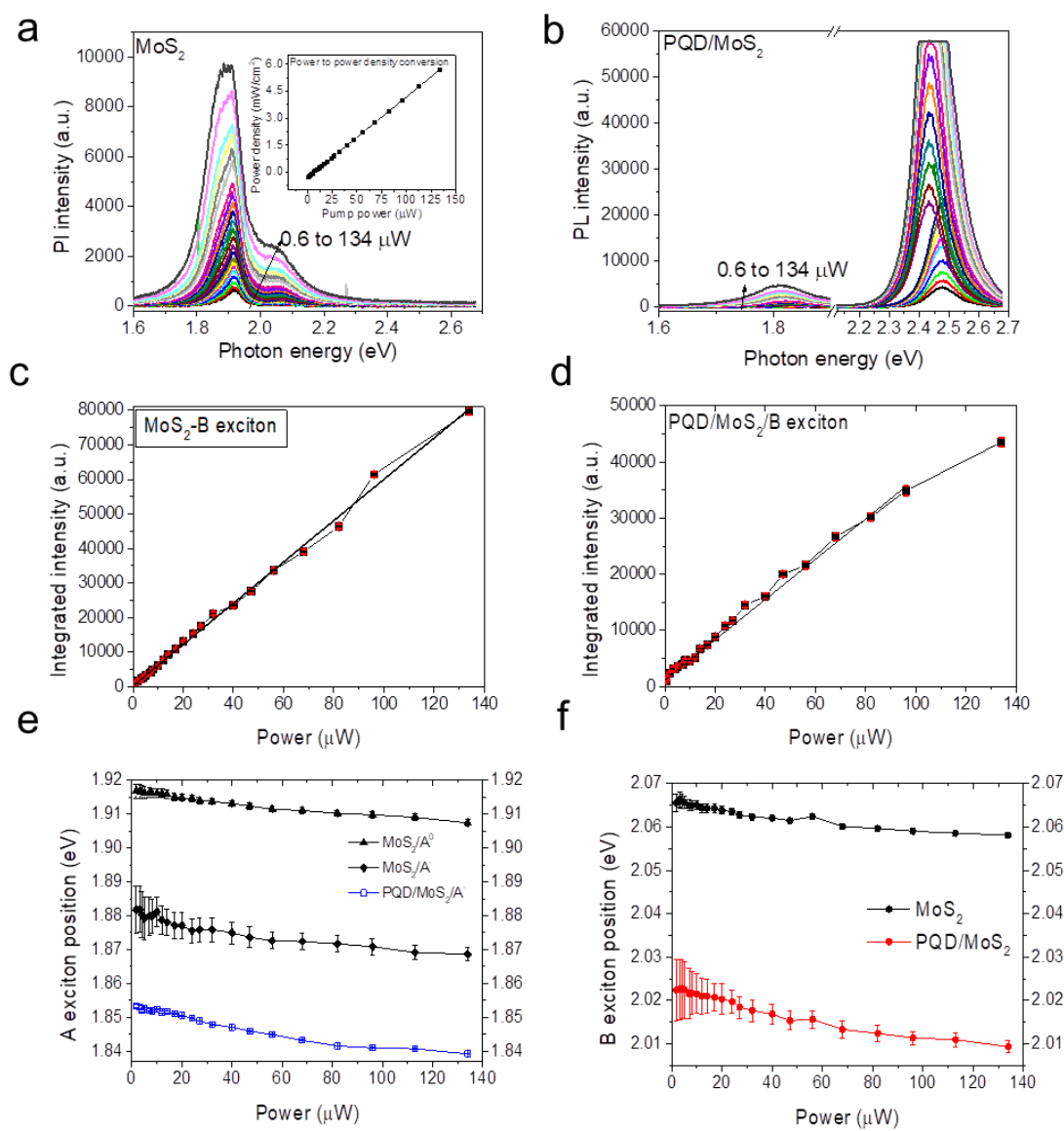

**Figure S12.** PL analysis of MoS$_2$ and PQD/MoS$_2$ heterostructure. (a) Power dependent PL spectra of MoS$_2$ and (b) PQD/MoS$_2$ with their integrated PL intensity plot of B exciton at various power in (c) and (d). The red line are linear fits of the integrated intensity. Inset in (a) shows the pump power converted to power density using a pinhole of 200 μm diameter. Lorentzian fitting of PL spectra produces maximum PL intensity peak (A), neutral exciton (A$^0$), trion (A$^-$) and B exciton which are summarized in (e) for A exciton and (f) for B exciton position. The excitation energy used in TA measurement was 3.062 eV (405 nm).

## S13. Fluence and kinetics of MoS$_2$ at E$_{exc}$ = 3.062 eV (405 nm)

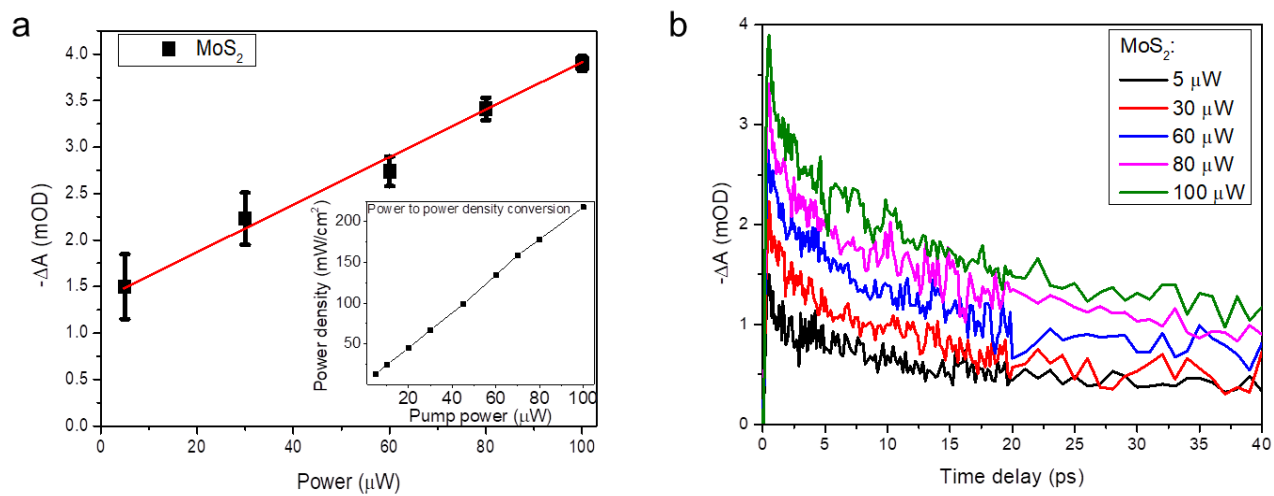

**Figure S13.** TA at different excitation fluences using pump excitation of 3.062 eV (405 nm). (a) Plot of bleach intensity versus fluence for MoS$_2$. The error bars are obtained from multiple spectra scanning. Inset shows the pump power converted to power density using a pinhole of 200 μm diameter. (b) TA kinetics of MoS$_2$ at various pump fluence obtained at B exciton position.

## S14. TA spectra analysis at $E_{exc}$ = 3.062 eV (405 nm)

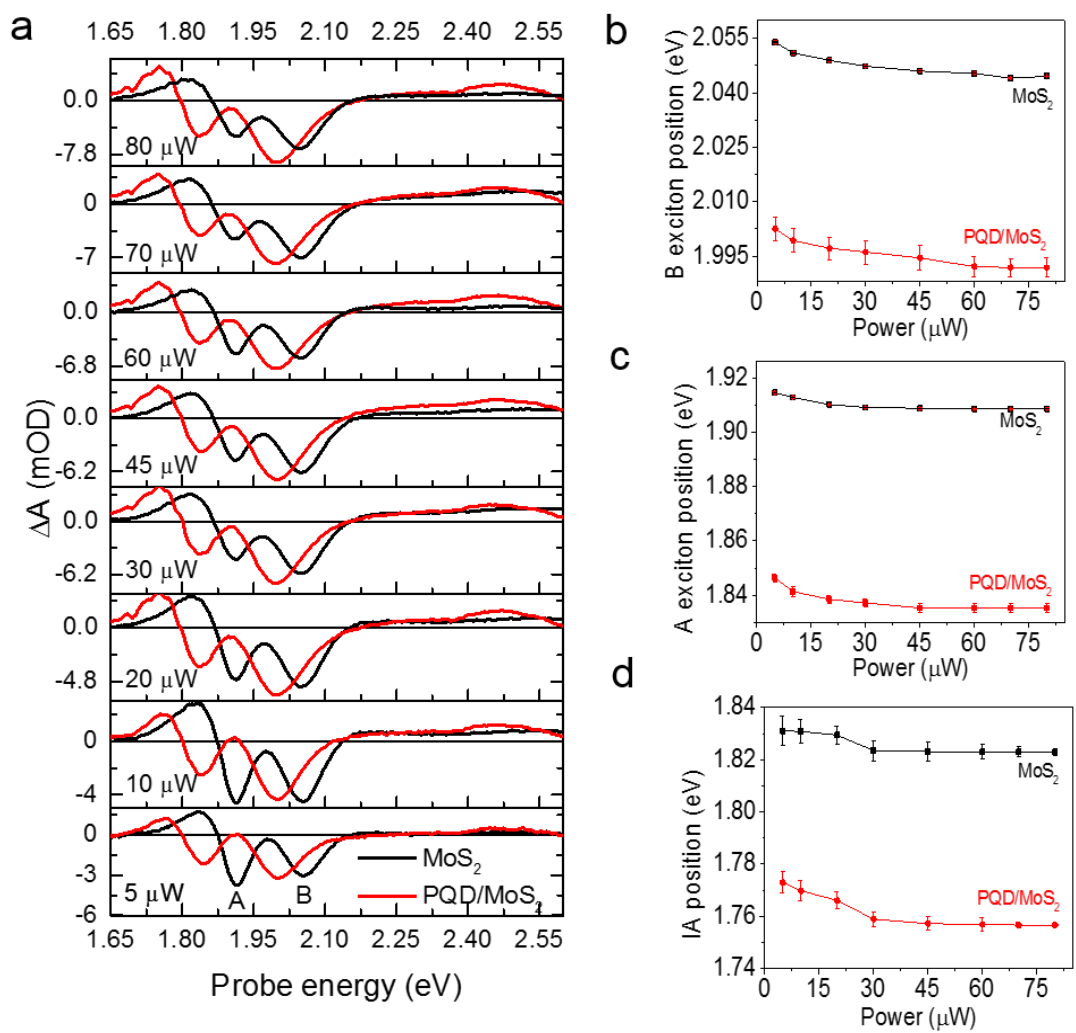

**Figure S14.** TA spectra of $MoS_2$ and PQD/$MoS_2$ heterostructure with 3.06 eV (405 nm) pump excitation. (a) Power dependent TA spectra of $MoS_2$ and PQD/$MoS_2$ heterostructure showing exciton bleach and induced absorption at 0.41 ps. Energy level position of (b) B exciton and (c) A exciton and (d) induced absorption (IA) position with different pump power. The error bars are estimated from the fluctuations in the TA intensity at multiple spectra scanning.

## S15. PL and TA correlation analysis at $E_{exc}$ = 2.331 eV (532 nm)

Similar TA and PL measurement as in Figure 5 were carried out with excitation source of 2.331 eV (532 nm) which excites only the excitons in $MoS_2$. We also observed onset and saturation of BGR for both A and B exciton purely due to the charge transfer from PQD. The comparable $\Delta E_x$ for A/B exciton in TA (60/45 meV) and PL (61/43 meV) further suggests that electron transfer from PQD itself (without the need of bandgaps excitation of PQD and $MoS_2$) can induce BGR in PQD/$MoS_2$ heterostructure.

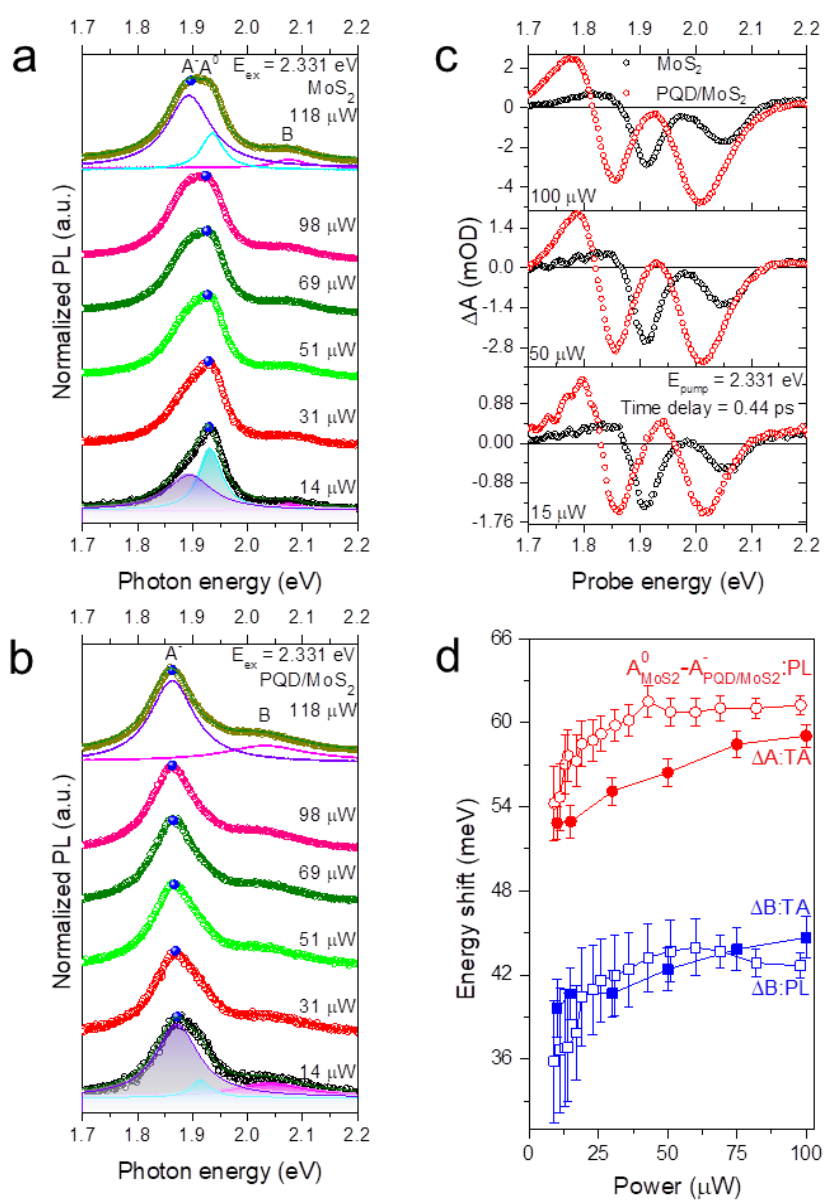

**Figure S15.** Correlation of emission and absorption process in BGR with 2.331 eV (532 nm) pump excitation. (a) Normalized PL of PQD/$MoS_2$ together with pristine $MoS_2$ in (b) at various laser power. The blue solid dots show the spectral shift of neutral exciton peak intensity in $MoS_2$ due to excitation density and electron transfer from PQD. (c) TA spectra of $MoS_2$ and PQD/$MoS_2$ at various pump power. Using excitation source which only excites $MoS_2$ excitons, redshift of exciton energy position can be observed in both PQD/$MoS_2$ PL and TA measurement. This redshift of exciton position from PL and TA within the error range obtained from fitting of TA spectra were plotted with their respective pump power in (d). The correlation of exciton position obtained from the absorption and emission processes shows that charge transfer doping from PQD due to band bending can also induce BGR in $MoS_2$.

## S16. TA and PL spectra and kinetic analysis at $E_{exc}$ = 2.331 eV (532 nm)

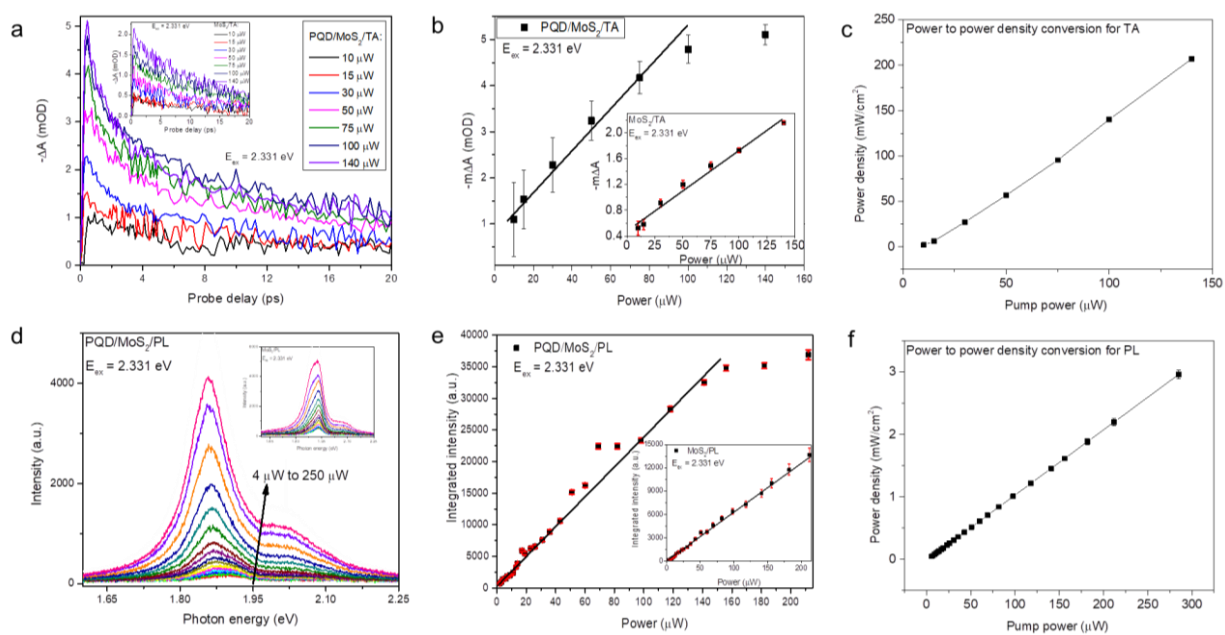

**Figure S16.** TA and PL analysis of $MoS_2$ and $PQD/MoS_2$ with of 2.331 eV (532 nm) excitation source. (a) TA kinetics obtained at B exciton position of $PQD/MoS_2$ at various pump power. Inset shows the same plot for pristine $MoS_2$. (b) Plot of TA bleach intensity of $PQD/MoS_2$ B exciton versus pump power. Inset shows the same plot for $MoS_2$. The red lines are linear fits of the bleach intensities and the error bars are estimated from the fluctuations in the TA intensity at various scanning. (c) The pump power to power density plot for TA measurement. Pinhole of diameter 200 µm was used for the conversion. (d) PL spectra of $PQD/MoS_2$ with pristine $MoS_2$ in the inset at various excitation power. (e) $PQD/MoS_2$ and $MoS_2$ (inset) integrated PL intensity plot of B exciton plotted against excitation power. The red lines are linear fit to the integrated intensities and the error bars are the fluctuations in the fitting of PL spectra. (f) Similar pump power to power density plot for PL measurement as in (c).

## S17. Linear absorption map and spectra of MoS$_2$ and PQD/MoS$_2$

For the practical implication of BGR, it is important to realize the exciton energy shift without the need of high photon energy. This is verified in our measurement from the linear absorption mapping of MoS$_2$ and PQD/MoS$_2$ heterostructure which shows A/B exciton shift of 18/25 meV.

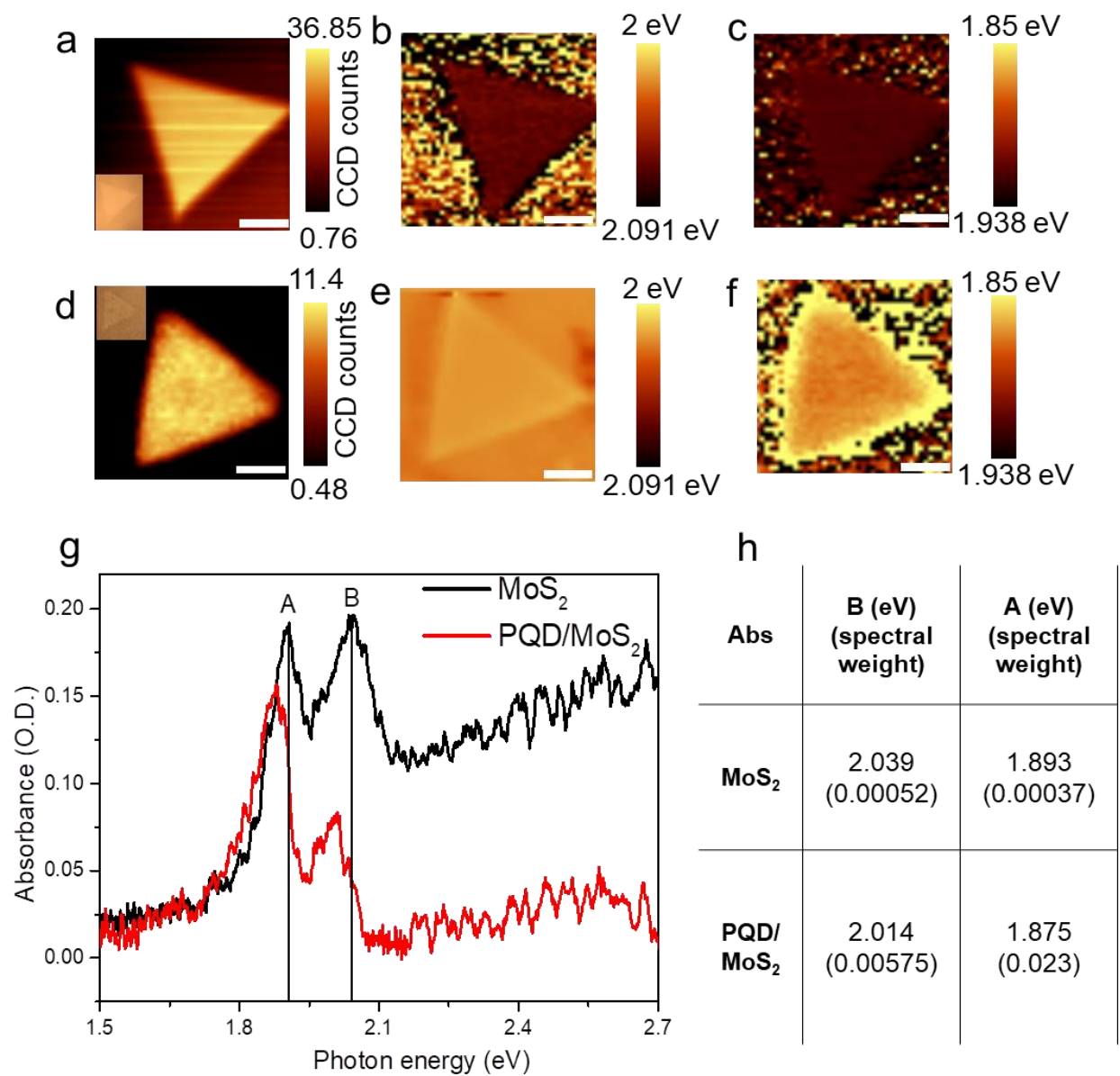

**Figure S17.** Confocal absorption mapping of MoS$_2$ and PQD/MoS$_2$. (a) Intensity mapping image of MoS$_2$ absorption spectra. Inset shows the optical image of MoS$_2$ flake. (b) MoS$_2$ B exciton and (c) A exciton mapping image with similar color scale bar for comparison. (d) Intensity mapping image of PQD/MoS$_2$ absorption spectra with (e) B exciton and (f) A exciton mapping image with similar color scale bar for comparison. (g) Averaged absorption spectra obtained from MoS$_2$ and PQD/MoS$_2$ absorption mapping. (h) Table summarizing the value of A and B exciton position in MoS$_2$ and PQD/MoS$_2$. The scale bar is 5 µm. The confocal absorption spectra imaging using a lab-made laser confocal microscopy system with a tungsten-halogen lamp was used to obtain absorption mapping and local spectra at a resolution of 1 µm diameter was detected by spectrometer.

**S18.  STM/S study of pristine and heterostructure MoS$_2$.**

The renormalization of exciton energy level in MoS$_2$ originates due to the change in exciton binding energy and the renormalization of electronic band states in MoS$_2$, giving rise to BGR. Hence to confirm the redshift of exciton energy level from the charge transfer mediated BGR in PQD/MoS$_2$ heterostructure, we performed the scanning tunneling microscopy and spectroscopy (STM/S) in pristine and PQD/MoS$_2$ heterostructure. Monolayer MoS$_2$ were transferred onto Si substrate after removing the SiO$_2$ layer using HF. Figure S18a shows the STM topography of 400 nm x 400 nm monolayer MoS$_2$ with FFT showing the sixfold symmetry of MoS$_2$ surface in the inset. Few bright spots in the image are attributed to the residual PMMA (during large area CVD MoS$_2$ transfer) while rest of the region shows clean MoS$_2$ surface. Similarly, Figure S18b shows the STM topography of 400 nm x 400 nm PQD/MoS$_2$ (Figure S18b) heterostructure with large portion of PQD islands (possibly due to the aggregation of nanocrystals) along with isolated nanocrystals and the pristine MoS$_2$ region. The FFT (see inset) obtained from the blue square region showing the sixfold symmetry indicates the existence of monolayer MoS$_2$ surface below the PQD islands.

Figure S18c shows the typical d$I$/d$V$ curve of pristine and heterostructure PQD/MoS$_2$ regions, showing the renormalization of the electronic band states in heterostructure as compared to pristine MoS$_2$. To extract the bandgap and other related electronic properties like conduction band minimum (CBM) and valence band maximum (VBM) from each d$I$/d$V$ curve, we followed the procedure as previously reported.[4] More than 80 d$I$/d$V$ curves recorded from different positions in Figure S18a and S18b, respectively, were used for this analysis (Figure S18d and S19). Figure S18d shows the statistically averaged CBM of the pristine and the PQD/MoS$_2$ heterostructure. Here we defined the Fermi level shift as the difference between the CBM position of the pristine MoS$_2$ and PQD/MoS$_2$ heterostructure which shows a value of around ~36 meV. This Fermi level shift also confirms the n-doping in MoS$_2$ due to electron transfer from PQDs to MoS$_2$ in the heterostructure. This n-doping effect thus lowers the CBM of PQD/MoS$_2$ heterostructure due to the band filling in MoS$_2$. Furthermore, lowering of the CBM can result in the reduction of exciton binding energy as well as the excitonic bandgap, as the excitonic band states in TMDs are bound to the CBM through the exciton binding energy.[5,6]

Based on the d$I$/d$V$ curves recorded from different positions, the CBM, VBM and the electronic bandgap are summarized in the histrogram plot in Figure S19 for MoS$_2$ and PQD/MoS$_2$ heterostructure. As compared to the lowering of the CBM, the VBM have a downshift and the electronic bandgap have a negligible shift in PQD/MoS$_2$ heteorstructure. It was reported that for MoS$_2$ transferred onto Si substrate, VBM is known to have larger fluctuations than CBM due to the local strain of the substrate (native SiO$_2$ in Si surface)[7] and additionally the trap states in SiO$_2$ can also produce local charge density variation in MoS$_2$.[8] These effects thus minimizes the overall bandgap reduction in our PQD/MoS$_2$ heterostructure. Moreover, the shrinkage of the CBM in PQD/MoS$_2$ strongly support our claim of exciton related electronic transitions originating from electrical bandgap renormalization.[4-6]

Additionally, the change in the CBM and the Fermi level shift (~36 meV) is consistent with the optical bandgap shift ($\Delta A/\Delta B$ = 18/25 meV) obtained from the linear absorption mapping of pristine and PQD/MoS$_2$ heterostructure, further confirming our charge transfer mediated BGR in PQD/MoS$_2$ heterostructure.

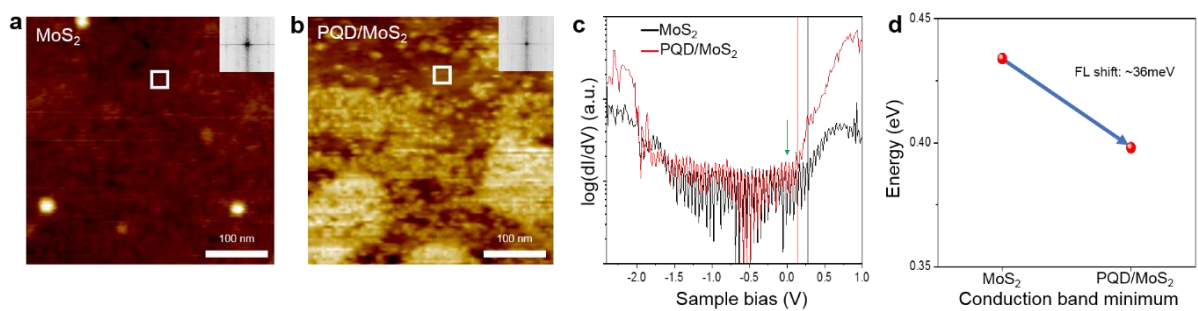

**Figure S18.** Scanning tunneling microscopy and spectroscopy (STM/STS) measurements of pristine MoS$_2$ and PQD/MoS$_2$ heterostructure. (a-b) STM images (V = 2.5V, I$_t$ = 500pA) of the pristine MoS$_2$ and PQD/MoS$_2$ heterostructure with insets showing FFT patterns obtained

from the white boxes, respectively. Both insets show the sixfold symmetry of $MoS_2$ surface. (c) Typical d*I*/d*V* curves in log-scale obtained from the pristine $MoS_2$ and $PQD/MoS_2$ heterostructure. Fermi level is at V = 0 and marked as a green arrow. Conduction band minimum (CBM) for each curve is marked as black and red line, respectively. (d) Averaged CBM of the pristine $MoS_2$ and $PQD/MoS_2$ heterostructure, respectively. The difference is ~36 meV, which is defined as the Fermi level shift.

## S19. STS analysis of pristine and heterostructure $MoS_2$.

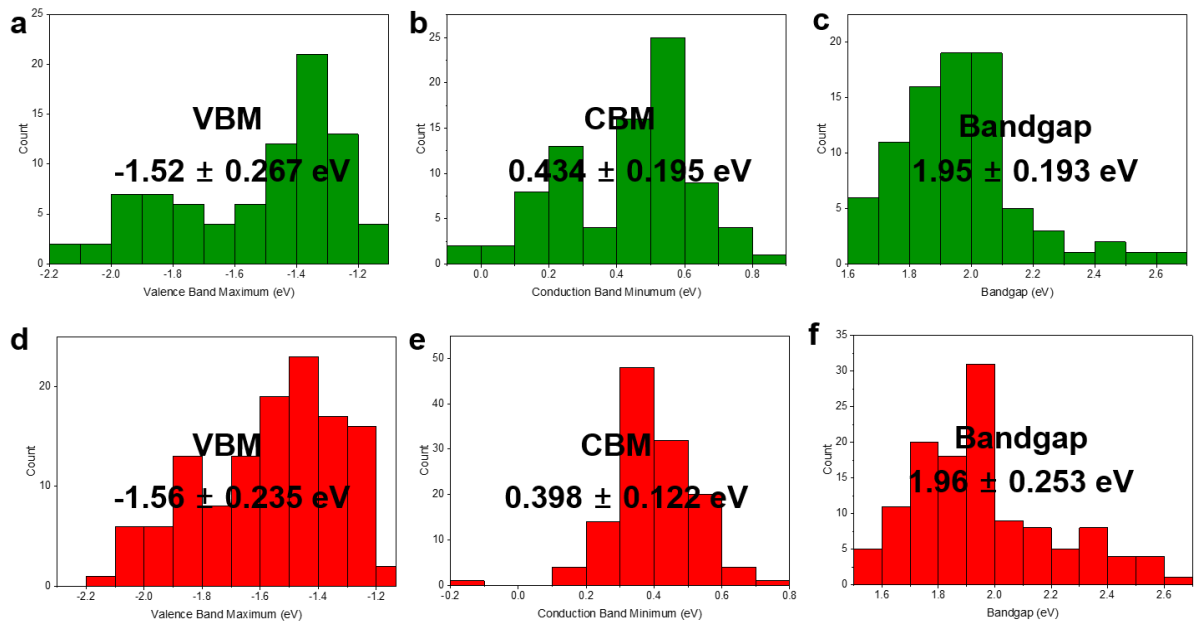

**Figure S19.** Statistical analysis on electronic properties of pristine $MoS_2$ and $PQD/MoS_2$ heterostructure. (a-c) Histograms of (a) VBM, (b) CBM and (c) Quasiparticle bandgap for the pristine $MoS_2$, respectively. (d-f) Histograms of (d) VBM, (e) CBM and (f) Quasiparticle bandgap for the $PQD/MoS_2$ heterostructure, respectively. For the pristine $MoS_2$ and $PQD/MoS_2$ heterostructure, more than 80 d*I*/d*V* curves collected randomly in the $MoS_2$ surface in Figure S18a and S18b, respectively, were processed to obtain the histogram plots.

## S20. *I-V* characteristics of MoS$_2$ and PQD/MoS$_2$ at dark and under illumination.

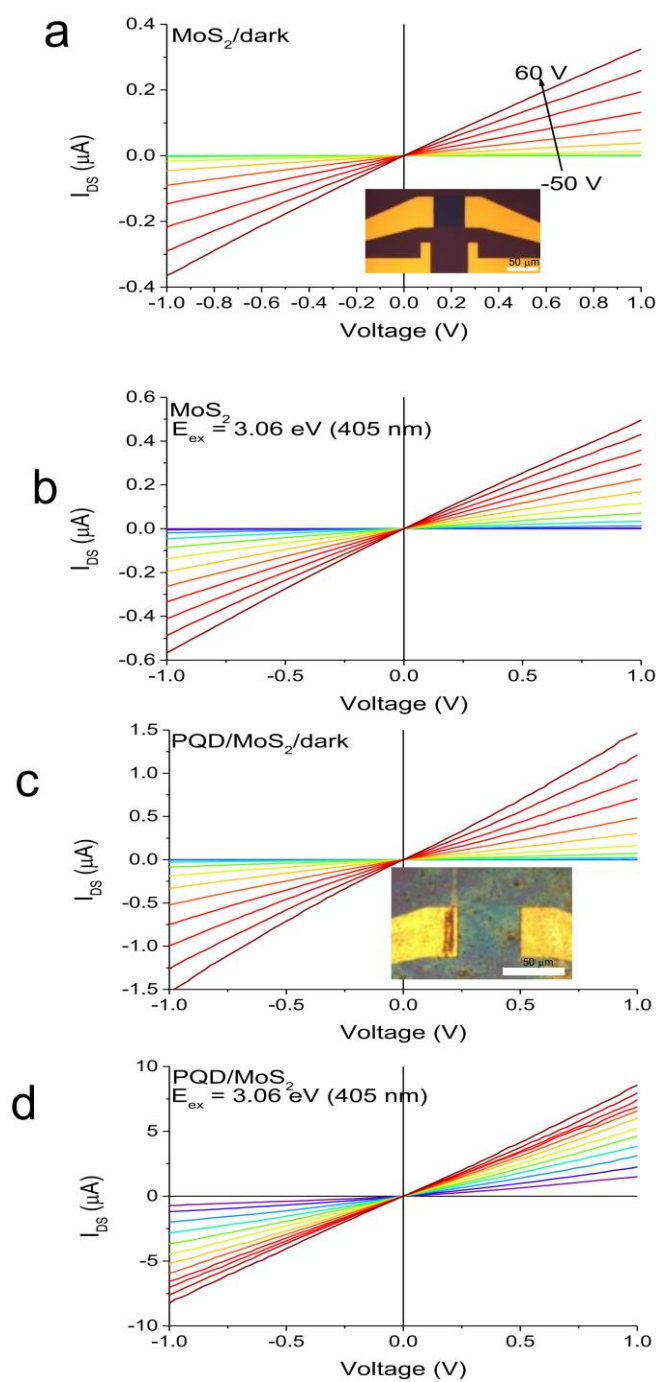

**Figure S20.** Dark and illumination *I-V* characteristics of MoS$_2$ and PQD/MoS$_2$. (a) $I_{DS}$-$V_{DS}$ plot of MoS$_2$ at dark and (b) under illumination. (c) $I_{DS}$-$V_{DS}$ plot of PQD/MoS$_2$ heterostructure at dark and (b) under illumination. The photo-illumination was carried out with 3.062 eV (405 nm) laser source at 134 µW. Inset in (a) and (c) are the device schematic of MoS$_2$ and PQD/MoS$_2$ with MoS$_2$ channel length of 50 x 100 µm$^2$.

## S21. Transfer characteristics of MoS₂ and PQD/MoS₂ for threshold voltage

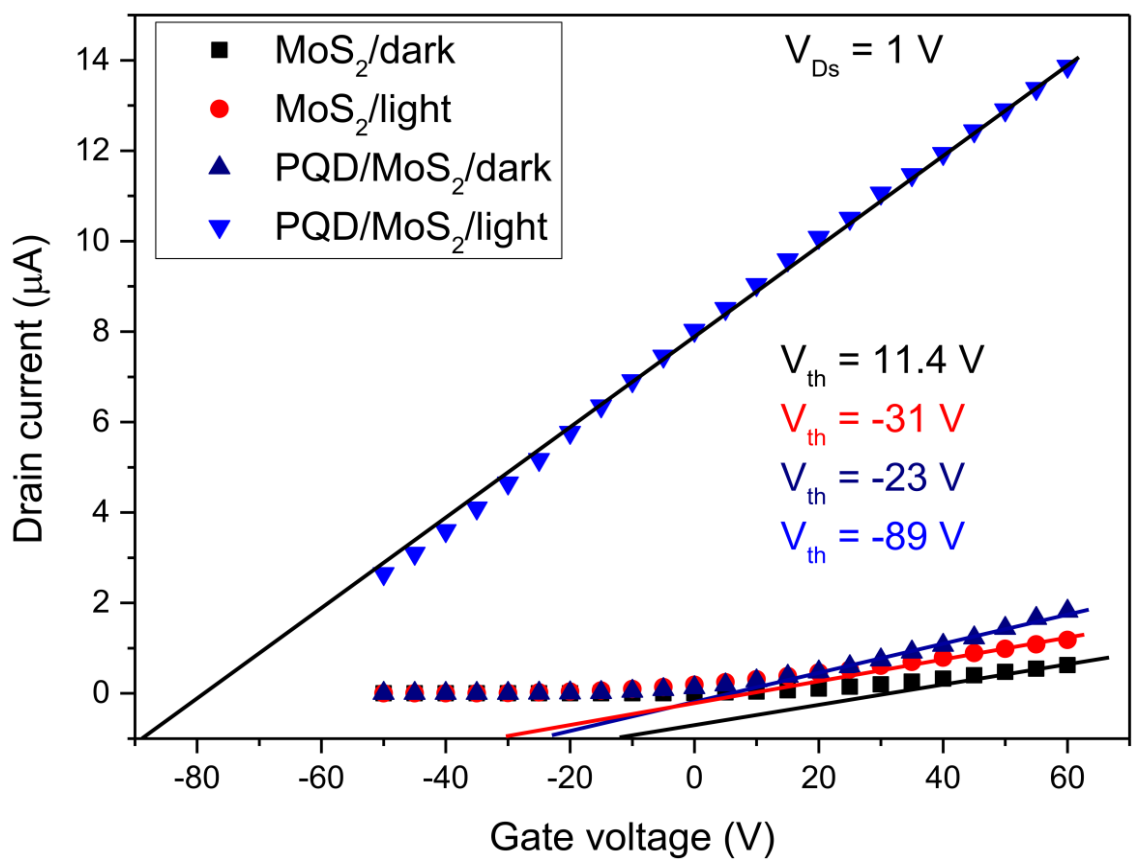

**Figure S21. Threshold voltage extraction.** Transfer characteristics of MoS$_2$ and PQD/MoS$_2$ at $V_{DS}$ = 1 V for dark and under illumination with excitation energy of 3.062 eV (405 nm) plotted. The linear current region extrapolated to the gate voltage in x-axis as shown by the solid lines were used for extracting the threshold voltage ($V_{th}$) values.

## S22. Thomas-Fermi Screening length

The screening of electric field by electrons in a solid can be quantatively analyzed from the Thomas-Fermi screening length (also known as Debye length) in a semiconductor given by:

$$L_{TF} = \sqrt{\varepsilon k_B T / e^2 N}$$

$k_B = 1.38 \times 10^{-23}$ $J/K$ is Boltzman constant,

$\varepsilon = \varepsilon_r \varepsilon_0 = 5.66 \times 10^{-11}$ $C^2/Nm$ is dielectric constant[9]

$T = 300$ $K$ is temperature and

$e = 1.6 \times 10^{-19}$ $C$ is electronic charge

$N = \frac{n}{t}$ cm$^{-3}$ is the carrier density per unit volume considering the thickness of MoS$_2$ ($t = 7.2 \times 10^{-10}$ m) while with n is 2D carrier concentration in monolayer MoS$_2$.